%% file: main.tex
\documentclass[acmsmall,screen]{acmart}
\AtBeginDocument{%
  \providecommand\BibTeX{{%
    \normalfont B\kern-0.5em{\scshape i\kern-0.25em b}\kern-0.8em\TeX}}}

\setcopyright{acmcopyright}
\copyrightyear{2022}
\acmYear{2022}
\acmDOI{10.1145/1122445.1122456}

\acmJournal{TiiS}
\acmVolume{37}
\acmNumber{4}
\acmArticle{111}
\acmMonth{8}

\usepackage{subcaption}
\usepackage{wrapfig}
\usepackage{color}

\usepackage{amsmath}
\usepackage{booktabs}
\usepackage{multirow}
\usepackage{rotating}
\usepackage{dblfloatfix} 
\usepackage{adjustbox}
\usepackage{graphicx}
\usepackage{lscape}

\DeclareMathOperator*{\argmax}{argmax}

\usepackage{soul}

\begin{document}

\title{GRAFS: Graphical Faceted Search System to Support Conceptual Understanding in Exploratory Search}

\author{Mengtian Guo}
\email{mtguo@email.unc.edu}
\affiliation{%
  \institution{University of North Carolina at Chapel Hill}
  \city{Chapel Hill}
  \state{North Carolina}
  \country{USA}
}
\author{Zhilan Zhou}
\email{zzl@live.unc.edu}
\affiliation{%
  \institution{University of North Carolina at Chapel Hill}
  \city{Chapel Hill}
  \state{North Carolina}
  \country{USA}
}
\author{David Gotz}
\email{gotz@unc.edu}
\affiliation{%
  \institution{University of North Carolina at Chapel Hill}
  \city{Chapel Hill}
  \state{North Carolina}
  \country{USA}
}
\author{Yue Wang}
\email{wangyue@unc.edu}
\affiliation{%
  \institution{University of North Carolina at Chapel Hill}
  \city{Chapel Hill}
  \state{North Carolina}
  \country{USA}
}

\input{tex/abstract}

\maketitle

\input{tex/introduction}

\input{tex/related_work}
\input{tex/method}

\input{tex/experiment}
\input{tex/results}

\input{tex/discussion}
\input{tex/conclusion}

\bibliographystyle{ACM-Reference-Format}
\bibliography{main}

\end{document}

%% file: tex/abstract.tex
\begin{abstract}

When people search for information about a new topic within large document collections, they implicitly construct a mental model of the unfamiliar information space to represent what they currently know and guide their exploration into the unknown. Building this mental model can be challenging as it requires not only finding relevant documents, but also synthesizing important concepts and the relationships that connect those concepts both within and across documents. This paper describes a novel interactive approach designed to help users construct a mental model of an unfamiliar information space during exploratory search. We propose a new semantic search system to organize and visualize important concepts and their relations for a set of search results. A user study ($n=20$) was conducted to compare the proposed approach against a baseline faceted search system on exploratory literature search tasks. Experimental results show that the proposed approach is more effective in helping users recognize relationships between key concepts, leading to a more sophisticated understanding of the search topic while maintaining  similar functionality and usability as a faceted search system.

\end{abstract}

%% file: tex/introduction.tex
\section{Introduction}

Exploratory search tasks are common among inquisitive users. Students explore scientific literature to gain knowledge about a topic. Health professionals synthesize medical literature to systematically assess treatments and associated outcomes for a disease. Journalists analyze news articles to link separate events into a coherent story. Intelligence analysts examine case reports to connect disparate evidence that suggests a potential threat. In all these tasks, searchers start with a complex information problem, yet a lack of understanding of the information space~\cite{white2009exploratory}. Such an understanding,  also called a ``mental model'' or ``schema'' in sensemaking literature~\cite{pirolli2005sensemaking}, may include key concepts or aspects that are important to the topic under investigation, and a rough understanding of how these concepts may relate to one another in the context of the topic. The mental model will co-evolve with the exploratory search process: it is updated as more information is encountered during search, and the perceived incompleteness of the model inspires further search activities~\cite{belkin1980anomalous}. 
Since such a mental model  plays a crucial role in exploratory search, it is desirable for systems to assist users in constructing a mental model at the beginning of an exploratory search task.

\begin{figure}[h!]
\centering
\begin{subfigure}[t]{0.55\textwidth}
  \includegraphics[width=\linewidth]{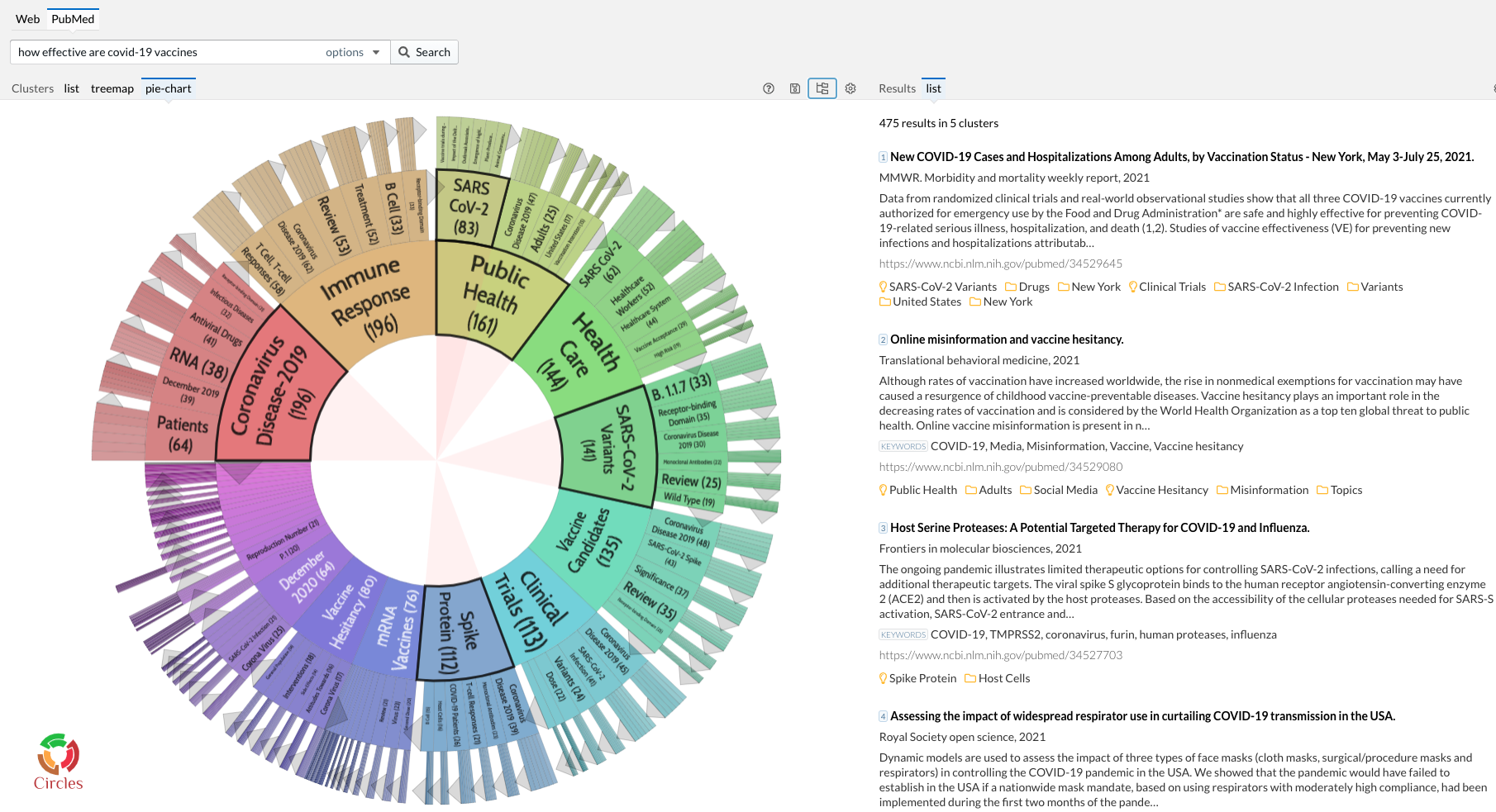}
  \caption{Carrot2}
  \label{fig:current-tools-a}
\end{subfigure}
\begin{subfigure}[t]{0.55\textwidth}
  \includegraphics[width=\linewidth]{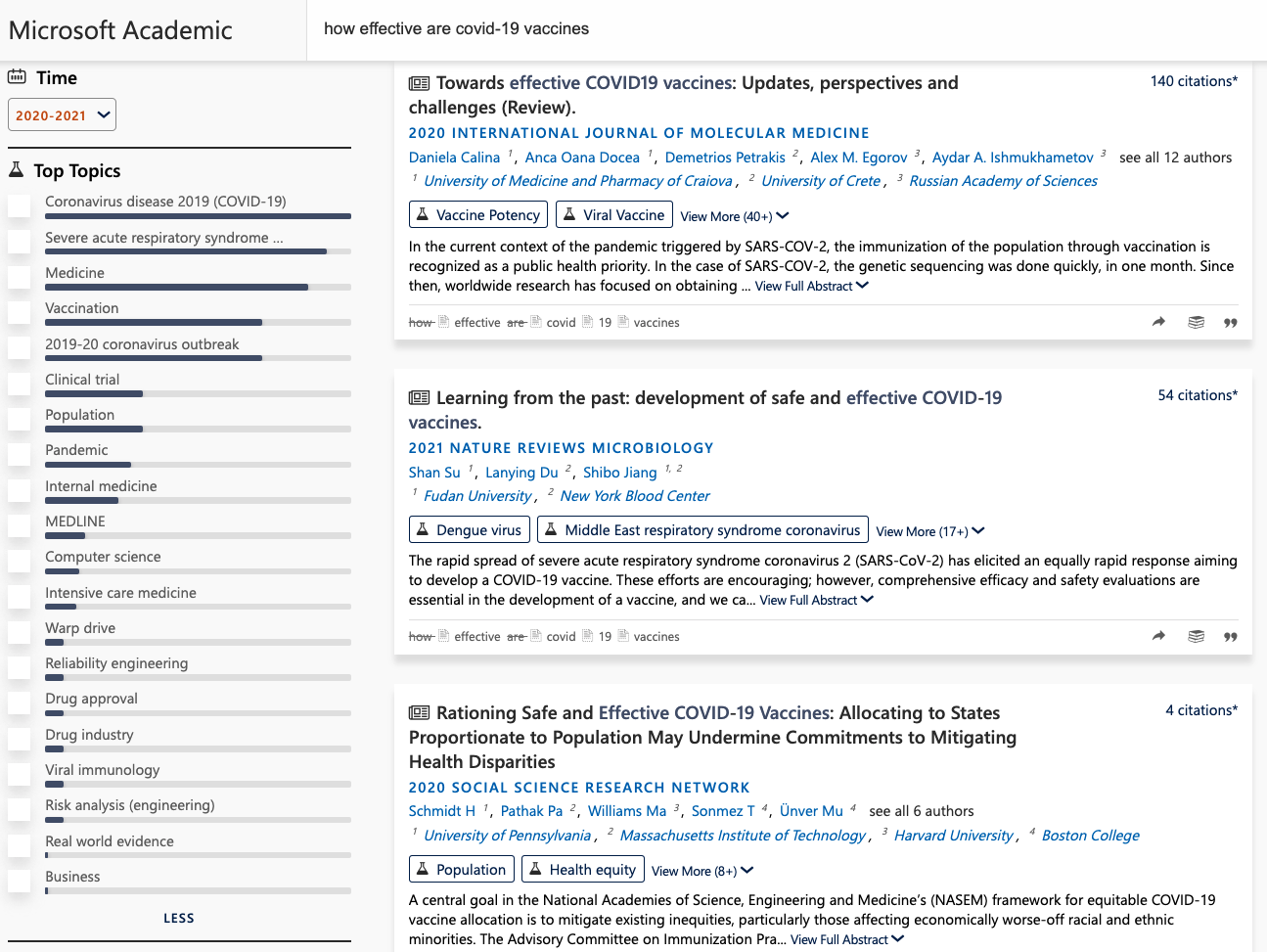}
  \caption{Microsoft Academic}
  \label{fig:current-tools-b}
\end{subfigure}
\begin{subfigure}[t]{0.56\textwidth}
  \includegraphics[width=\linewidth]{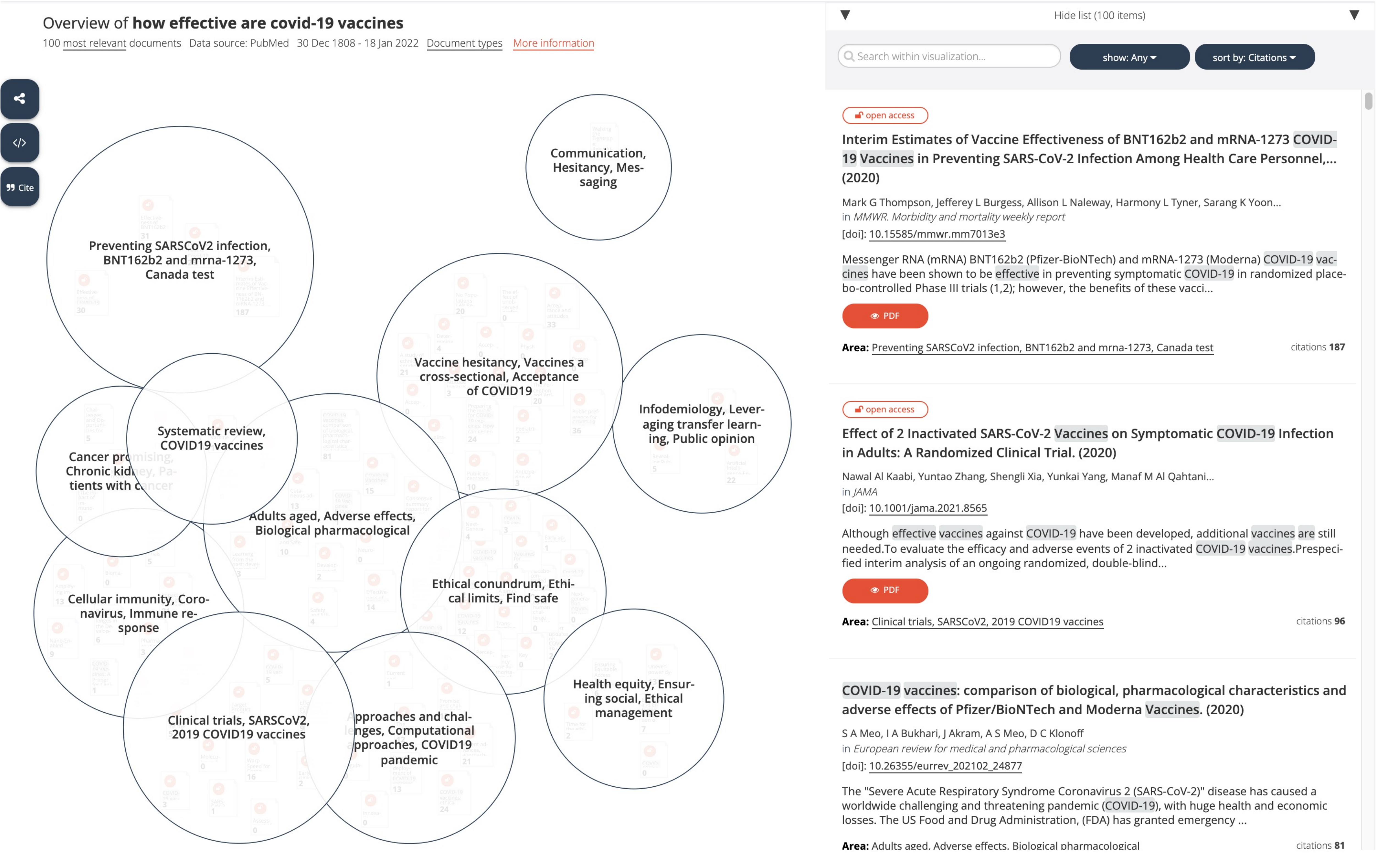}
  \caption{Open Knowledge Map}
  \label{fig:current-tools-c}
\end{subfigure}
\Description[Screenshots of Existing Tools]{Six screenshots of existing search tools that show semantic annotations in the search interface}
\caption{Examples of current search systems that show semantic annotations as different interface elements. 
(a) Carrot2~\cite{osinski2004lingo} infer a hierarchical cluster structure over search results and visualize the hierarchy in a pie-chart or treemap.
(b) Microsoft Academic~\cite{wang2020microsoft} annotates each article with research topics and uses them as faceted filters on the left side of the results list.
(c) Open Knowledge Maps~\cite{kraker2016open} organizes results into semantic clusters and uses spatial layout to show semantic relationship between clusters. Accessed in January 2022.}
\label{fig:current-tools}
\end{figure}

Existing search engines offer varying degrees of support for exploratory searchers to build a mental model of the information space. The classical  list presentation (e.g., Google) provides minimal support (if any) as the list of result snippets may or may not contain important concepts or their relations. Document clustering search engines aim to group search results into clusters and automatically summarize clusters using keywords (Figure \ref{fig:current-tools-a}). They attempt to organize search results into conceptual groups, which may help users build a mental model.  However, algorithm-generated document clusters and cluster-labels are not guaranteed to be meaningful even in the eyes of experts. The potential unintelligibility adds to the burden of a user who is exposed to an unfamiliar information space for the first time. 

As semantically annotated documents become available through manual or automated annotation approaches, search engines can begin to surface meaningful semantic concepts to aid exploratory searchers. For example, a facet panel groups semantic concepts into different categories, i.e., facets (Figure \ref{fig:current-tools-b}). It provides the user with an overview of important concepts in the search results, as well as the capability to filter results by specific concepts. However, a facet panel cannot effectively communicate how these concepts \emph{relate} to one another. It only implicitly reveals the connections between a selected concept and other unselected concepts if it shows the volume of search results associated with each concept (e.g., bars of different lengths in the facet panel in Figure \ref{fig:current-tools-b}). 

Other search interfaces visualize concepts in a 2D layout, using spatial proximity as a metaphor to communicate conceptual relationships (Figure \ref{fig:current-tools-c}). This works well if the visualized concepts are few and their relationships are sparse. However, in many search scenarios, the results may contain a large number of concepts that are densely related to each other. In those scenarios, the visualization tends to clutter the interface with many concepts and relations, while also inevitably losing information due to the projection of the complex network of concepts into a two-dimensional representation.


In our work, we envision an intelligent search system that can help exploratory searchers discover key concepts and conceptual relationships in search results through a simple, minimally cluttered interface. 
This is a meaningful and challenging problem. On the one hand, seeing how key concepts interplay with each other helps the searcher build a more complete mental model of the information space, which can enable more informed and fruitful investigation. On the other hand, the number of relationships between key concepts can be too large, and presenting them may easily overwhelm the user who just started the exploration. An ideal system would start by revealing just enough key concepts and relations needed to construct an initial mental model, and progressively nudge the searcher towards an increasingly nuanced understanding of the information space over the course of search. In this paper, we begin to address this problem through three key contributions.
\begin{itemize}
    \item First, we design a new interface to communicate key aspects of a computationally maintained knowledge subgraph to users during interactive search tasks. The design objectives include: (1) reveal relationships between concepts without cluttering the interface; (2) leverage user's familiarity with existing search interface elements such as a main result listing and a facet panel; and (3) support a mixed-initiative approach where users can make changes to the subgraph as needed. This results in the \underline{Gra}phical \underline{F}aceted \underline{S}earch system, or GRAFS. In GRAFS, a knowledge subgraph is embedded in a familiar faceted search-like interface to help searchers construct a basic understanding of the information space and encourage them to dig deeper. 
    \item Second, an exploratory searcher's need for a mental model inspires us to formulate a new data model – an initial knowledge subgraph for exploratory search. The goal of this data model is to initiate the construction of a mental model in the searcher's mind. We propose that the subgraph shall meet the following criteria: (1) be relevant to the search context; (2) have wide coverage of the information space to stimulate learning; and (3) be efficiently computable to support real-time interaction. We propose an efficient algorithm for extracting such a knowledge subgraph given semantic search results. 
    \item Third, we provide results from a user study conducted to evaluate the proposed search interface in the context of medical literature exploration tasks. Compared to a classical faceted search interface, GRAFS  helps users better identify relations between concepts and develop a deeper understanding of search topics. At the same time, GRAFS maintains the original functionality of a faceted search system with which users are familiar.
\end{itemize}

%% file: tex/related_work.tex
\section{Related Work}

The proposed search system aims to assist users in exploratory search tasks.  Its frontend interface leverages faceted filters, result clustering, and information visualization. Its backend data model computes and maintains a knowledge subgraph from semantic search results and user interaction. As such, this work is related to multiple lines of previous work in interactive and intelligent systems as we discuss below.

\subsection{Exploratory Search}
Exploratory search happens when a user has a complex information problem but insufficient knowledge to clearly express the information need~\cite{white2009exploratory}. Answers to such problems often cannot be found in any one document, but instead has to be synthesized over many documents retrieved by a series of revised queries. Exploratory searchers alternate between two modes: learning and investigation~\cite{marchionini2006exploratory}. In the learning mode, a searcher tries to make sense of important aspects encountered in the search results and how these aspects relate to one another, i.e., construct a mental model of the information space. Realizing missing pieces in the current mental model, a searcher will enter the investigation mode to collect more information to fill in the gap, leading to the next round of learning. Common web search engines, such as Google and Bing, are primarily designed for fact retrieval, with limited explicit support for learning activities.  More elaborate interactive features have been proposed to support exploratory search~\cite{white2005exploratory}, including facet filters~\cite{yee2003faceted,kules2008users}, result clustering~\cite{zamir1999grouper,carpineto2009survey}, and information visualization~\cite{hoeber2018information,dowling2019interactive}.  Our goal in this work is to design search systems that provide better support for exploratory search, especially for the discovery of key concepts and their relations in the learning stage. 

\subsection{Faceted Filtering and Search Result Clustering}
Faceted search systems are commonly used in domains where documents are associated with rich metadata. These search systems group different dimensions of metadata into facets which allow a user to slice and dice search results along different facets and facet-values~\cite{tunkelang2009faceted}. The list of relevant facet-values, sometimes each annotated with a corresponding number of associated results, naturally provides an overview of the information space. These powerful capabilities make faceted search systems amenable to exploratory tasks~\cite{white2009exploratory}.  Faceted search systems have successful applications in digital libraries~\cite{fagan2010usability} and e-commerce websites~\cite{niu2019understanding}, where rich metadata have traditionally been manually assigned to each document. With the help of supervised machine learning and natural language processing techniques, metadata from a pre-defined knowledge graph or ontology can be automatically assigned to documents~\cite{shen2014entity}.

Search result clustering systems also aim to organize a large collection of results into subgroups~\cite{carpineto2009survey}. Each subgroup is often assigned a word or phrase label generated by the clustering algorithm. This approach can be applied to arbitrary textual search results without manual annotation efforts, as cluster labels and document groupings are automatically generated by text clustering algorithms. However, algorithm-generated clusters and cluster labels can sometimes be difficult to interpret and can lead to user confusion~\cite{hearst2006clustering}.

The approach outlined in this paper draws on the strengths of both faceted search and clustering. We augment the traditional facet panel with relationship links that are interactively surfaced between semantic concepts (facet-values). Moreover, we cluster key concepts such that semantically related concepts are represented closely in the facet panel, nudging users to see connections among related (and therefore nearby) concepts.

\subsection{Text Search Result Visualization}
Information visualization provides powerful approaches for users to view, search, manipulate, and reason about complex textual information through graphical representations. Information retrieval systems have a long history in employing visualizations to help users obtain an overview of search results~\cite{hearst2009search}. A common approach is to project documents, keywords, and concepts as objects on a two-dimensional canvas or nodes in a network, such that spatially close objects are semantically related~\cite{cao2010facetatlas, van2010software, nocaj2012organizing, kraker2016open, dowling2019interactive, husain2021multi}.
When additional document metadata are available, search results can also be visualized as with additional organization such as topical facets~\cite{wang2020microsoft}, timelines~\cite{lu2017visual}, geographic maps~\cite{teitler2008newsstand}, and knowledge graphs~\cite{sarrafzadeh2017improving}.
However, visualizing nominal, high-dimensional textual information in a low-dimensional space with limited screen resolution will necessarily lead to loss of information. Therefore, to make full sense of visualized search results, users often still need to read associated documents~\cite{hearst2009search}.

In the approach outlined in this paper, we visualize not retrieved documents but important concepts and relations in retrieved documents, which constitute a knowledge graph. Recognizing that an exploratory searcher is typically not familiar with all concepts and their relations until after reading associated texts, we keep the visualization simple and progressive.  This approach introduces additional detail as an exploratory search task unfolds as opposed to taking up large screen area or visualizing all details of the subgraph upfront.

\subsection{Knowledge Graph in Search Systems}
Knowledge graphs, or semantic networks, play important roles in search systems. Semantic search technology relies on accurate recognition of concepts in queries and documents to achieve backend capabilities such as query intent understanding~\cite{blanco2015fast}, automatic query expansion~\cite{dalton2014entity}, improved result ranking~\cite{xiong2017explicit}, entity retrieval~\cite{balog2011query}, and direct answer retrieval~\cite{sun2015open}. In the frontend, semantic concepts afford new interactive browsing features in addition to document listings. These include per-result semantic tags~\cite{wei2019pubtator}, concept-based filtering~\cite{wang2020microsoft, soto2019thalia}, faceted navigation~\cite{soto2019thalia,hope2020scisight}, and conversational search~\cite{vakulenko2017conversational}. 

The work introduced in this paper focuses on enabling user interactions with a query-specific knowledge graph during exploratory search. 
Previous work in natural language processing has proposed various approaches to constructing a query-specific knowledge graph, with the primary goal of improving relevance ranking of documents or entities~\cite{dalton2013constructing, schuhmacher2015ranking, ensan2019relevance, dietz2019ent}. Although we also construct a query-specific knowledge graph in this work, the primary goal is for front-end presentation as a means to assist exploratory searchers whose goal is not necessarily to identify the most relevant documents or entities, but rather to make sense of an unfamiliar topic and corresponding search results with varying degrees of relevance.
Previous search interfaces either present a set of relevant concepts without surfacing their relations~\cite{ wang2020microsoft, soto2019thalia}, or use a large screen area to visualize concepts and relations (in which the main document listing area is substantially reduced, deviating from the familiar traditional search interface)~\cite{kraker2016open,sarrafzadeh2017improving, hope2020scisight}. 
In contrast, our design objective is to show relations between concepts while still maintaining key aspects of the traditional and effective document listing design common to most users' search experiences.

%% file: tex/method.tex
\section{Graphical Faceted Search}
\label{sec:Graphical Faceted Search}
In this section, we describe the design and implementation of the graphical faceted search (GRAFS) system. We start by motivating our research problem. We then divide the problem into two parts: a data representation problem and an interactive presentation problem. In each part, we formulate the problem, discuss the general solution framework, and describe our specific implementation of the solution.

\subsection{Challenge and Motivation}
\label{sec:Challenge and Motivation}

An exploratory searcher needs to obtain an overview of a large number of potentially relevant documents returned by her query. Instead of reading these documents one by one trying to build a conceptual understanding of the information space, a better strategy is to learn key concepts and relations that show up frequently throughout the search results. In fact, semantic search engines also generate and leverage such information in their backend~\cite{reinanda2020knowledge}.

\begin{figure}
\centering
\begin{subfigure}{.5\textwidth}
  \centering
  \includegraphics[width=.76\linewidth]{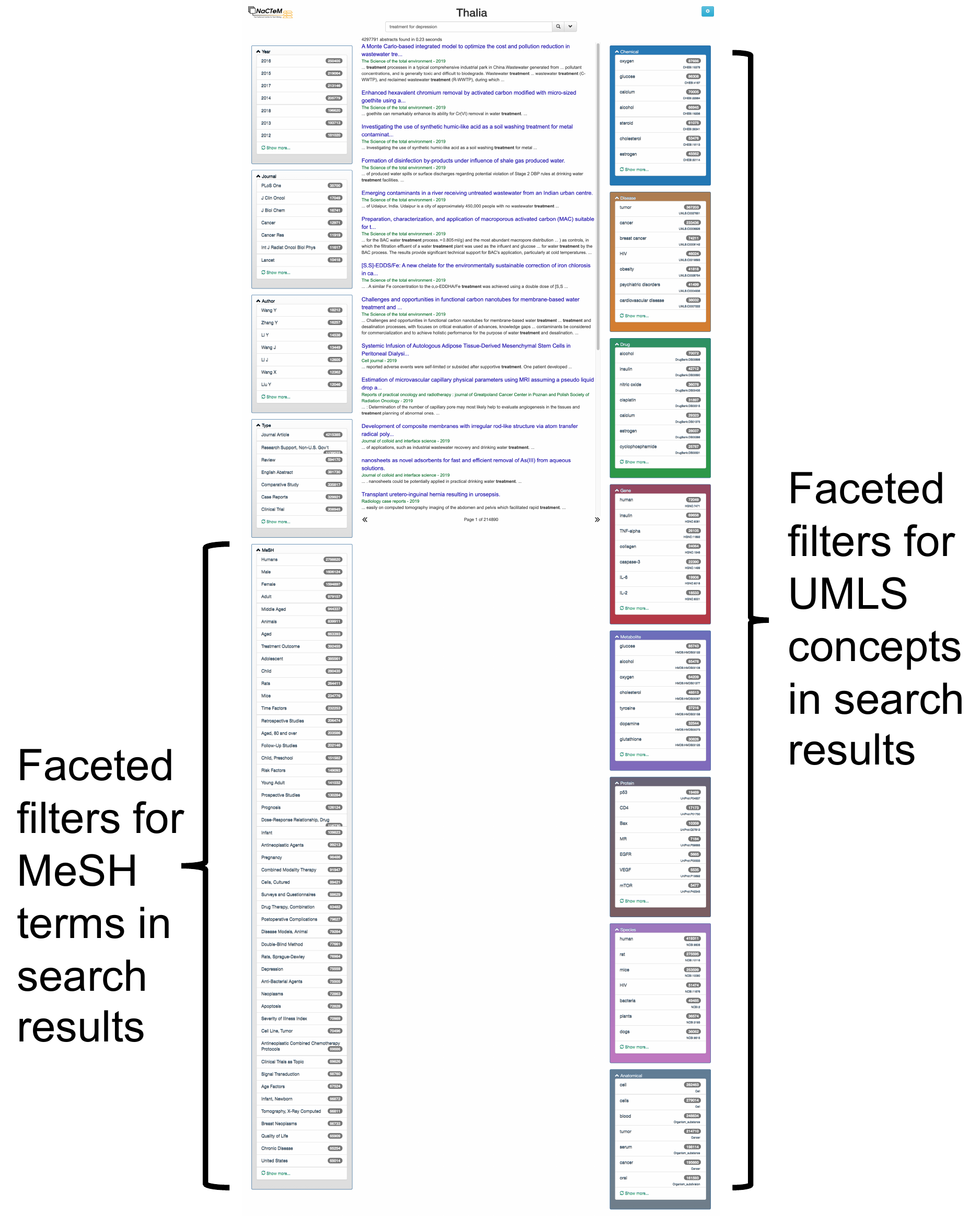}
  \caption{Faceted filters}
  \label{fig:challenge-facet}
\end{subfigure}%
\begin{subfigure}{.5\textwidth}
  \centering
  \includegraphics[width=.85\linewidth]{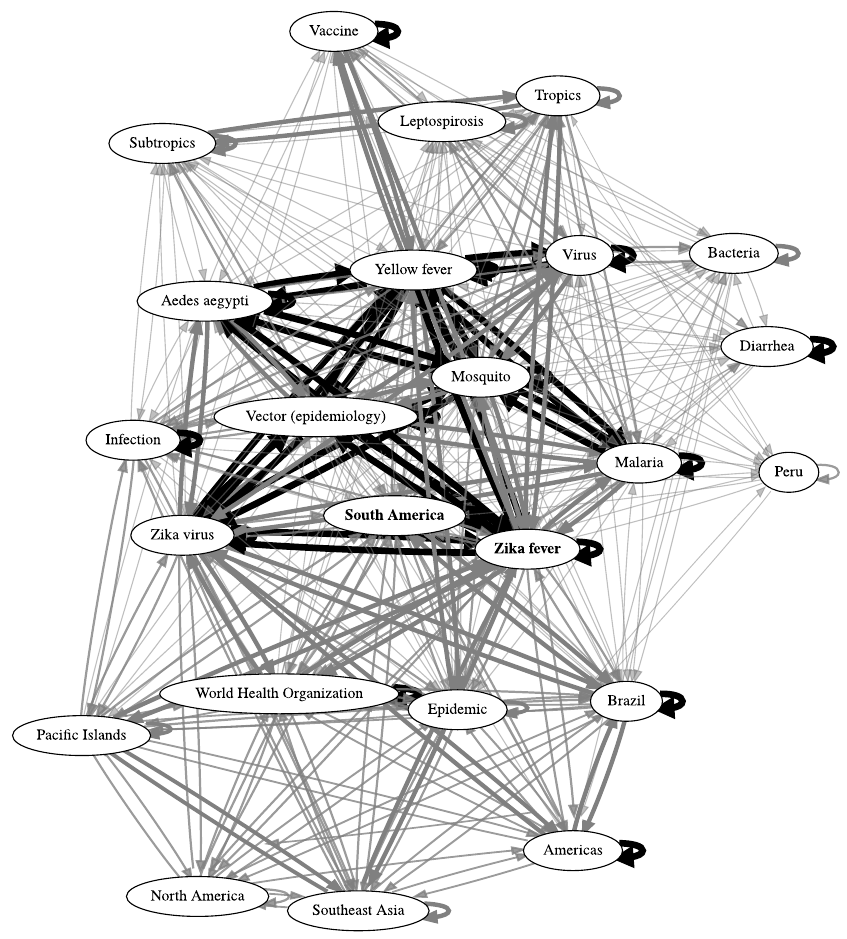}
  \caption{Knowledge subgraph visualization}
  \label{fig:challenge-vis}
\end{subfigure}
\caption{Challenges in presenting concepts and relations in search results. (a) A screenshot of Thalia search engine's result page for the query ``treatment for depression'' over PubMed abstracts~\cite{soto2019thalia}. Hundreds of MeSH terms and UMLS concepts mentioned in the search results are used as faceted filters. This leads to exceedingly long lists of faceted filters on two sides of the interface. (b) A visualization of the knowledge subgraph constructed for the query ``Zika fever'' on English Wikipedia (reproduced with author's permission) ~\cite{dietz2019ent}. Each node is a DBpedia concept mentioned in the search results. Two concepts are connected by an edge if they are mentioned in the same context in any search result. Showing all relationship links is visually overwhelming even after pruning the number of concepts down to two dozen. }
\label{fig:challenge}
\end{figure}

A semantic search engine applies entity recognition and linking algorithms to documents (at index time) and queries (at search time). This has become increasingly common thanks to the advances in natural language processing (NLP) and the availability of knowledge graphs  (also called standardized vocabularies or ontologies) in different domains. As a result, when a search engine returns a list of documents, it can simultaneously return a set of entities (and their relations) recognized in these documents. In other words, a semantic search not only returns a set of documents, but also a set of semantic units, a subgraph inside the global knowledge graph. On the one hand, this knowledge subgraph has great potential to help an exploratory searcher obtain a mental model of the information space in question. On the other hand, this knowledge subgraph can be enormous if it is shown in its raw form. It may consist of not only core concepts of the information space but also a large number of peripheral and even non-relevant concepts that happen to be mentioned in the result documents. Therefore, \textbf{it is challenging to allow exploratory searchers make use of a query-specific knowledge subgraph in a non-overwhelming manner.} 

To illustrate this challenge, consider NaCTeM's Thalia search system for PubMed abstracts~\cite{soto2019thalia}. The system uses NLP algorithms to extract UMLS concepts and MeSH terms from abstracts and use them as faceted filters on the result page. As shown in Figure \ref{fig:challenge-facet}, the query ``treatment for depression'' retrieves an exceedingly long list of concepts from search results, which resulted in faceted filters vertically spanning multiple screens! Even worse, top-ranked concepts in each semantic category (facet) are most frequent yet non-informative, such as  \emph{Human} in \emph{Gene/Species}, \emph{Cell} in \emph{Anatomical Parts}, and \emph{Oxygen} in the category \emph{Chemical}.

To reduce the complexity of the query-specific knowledge subgraph, researchers have proposed selection and ranking algorithms aiming to preserve only a set of core concepts~\cite{dalton2013constructing, schuhmacher2015ranking, dietz2019ent}. However, even after reducing the space of concepts, the number of possible relationships among concepts can still be large. Figure \ref{fig:challenge-vis} shows a knowledge subgraph produced by a state-of-the-art approach when searching Wikipedia articles using the query ``Zika fever''~\cite{dietz2019ent}. Despite having only 24 concepts, the subgraph is densely connected due to intricate relationships between these concepts in search results. Showing all relationship links upfront can easily overwhelm a user's mind.

The above challenge directly motivates our work. Below we describe our approach to this challenge, including interactive designs for presenting a knowledge subgraph for exploratory search, as well as computational methods for constructing and maintaining the underlying data model. As a specific use case, we construct and present knowledge subgraphs to medical literature searchers in an uncluttered and non-overwhelming manner. 



\subsection{Interaction Design and Data Model}

In this section, we first present our interaction design goals for GRAFS. We then design the underlying backend data model, i.e., a knowledge subgraph, that supports the frontend design goals. 

\subsubsection{Interaction Design Goals}
\label{sec:Interactive Design Goals}

Informed by the challenges of exploratory search described above, we propose the following design goals (DGs) for the interactive experience to be fulfilled by GRAFS.
\begin{itemize}
    \item [\textbf{DG1}] \textbf{Preserve the simplicity, familiarity, and capability of a search interface.} Search engine users are well educated to use the Google-like list presentation and faceted navigation interface. We aim to maintain the usability of the new search interface by leveraging the basic layout and functionality of existing systems.
    \item [\textbf{DG2}] \textbf{Show relations between concepts without cluttering the interface.}  Visualizing relations between concepts have great potential to help users explore the information space. However, high-density connections can easily clutter the interface and discourage learning. We aim to selectively expose these relations in a sparse, non-cluttering manner.
    \item [\textbf{DG3}] \textbf{Preserve human agency.} The visualization of concepts and relations, i.e., a knowledge subgraph, is meant to initiate the user's learning activities. 
    Users should have the ability to make adjustments to the knowledge subgraph based on their mental model or current interest during the sensemaking process.
\end{itemize}



\subsubsection{Data Model Formulation}

The above interaction design goals imply a data model that supports user interaction. The data model is a small-scale knowledge graph that contains the most informative concepts and relations embedded in a large number of search results. The goal of this data model is to inspire users to learn about these concepts and relations and form their own mental model of the information space, i.e., a parsimonious and evolving understanding of key concepts in the space and how they relate to one another. The data model aims to lead the user in constructing their mental model, but it is not the mental model itself. Below we formulate this data model.

A semantic search engine returns not only a ranked list of results for a given query but also concepts (or entities) and their relations within those results. Formally, an initial search query $q$ retrieves a list of documents $D_q = \{d_i\}_{i=1}^n$. Within each document $d_i$, the search system also recognizes a set of concepts $C_i \subset C$ and a set of relations $R_i \subset R$.\footnote{We use standard set-theoretic notation in this paper. ``A $\cup$ B'' means the union of sets $A$ and $B$. ``$A \cap B$'' is the intersection of sets $A$ and $B$. ``$e \in A$'' means element $e$ is a member of set $A$. ``$A \subset B$'' means $A$ is a subset of $B$. ``$A \setminus B$'' means the subset of $A$ that is not in $B$. ``$\varnothing$'' means the empty set. ``$\bigcup_{i=1}^n C_i$'' stands for the union of $n$ sets $C_1, C_2, \cdots, C_n$ \ . }
Here, $C$ is the set of all possible concepts in the global knowledge graph, and $R$ is the set of all possible relations between those concepts. Therefore, in addition to the list of documents $D_q$, the initial search query $q$ effectively retrieves a query-specific knowledge graph $G_q = (C_q, R_q)$, where $C_q = \bigcup_{i=1}^n C_i$ and $R_q = \bigcup_{i=1}^n R_i$. We assume that concepts and relations in $G_q$ have been pre-extracted by NLP algorithms in the search engine backend~\cite{dalton2013constructing, wolfe2017pocket, reinanda2020knowledge}. As described in Section \ref{sec:Challenge and Motivation}, $G_q$ often contains a large number (hundreds or even thousands) of concepts and relations. Directly presenting the entire graph $G_q$ to users will inevitably cause information overload. 

To support the above interaction design goals, we need to construct an \textbf{initial knowledge subgraph} $H_q \subset G_q$ that is most helpful at the initial stage of mental model construction. Just as a traditional search interface tackles information overload by showing a small number (e.g., 10) of the most relevant documents on the first page, we posit that the GRAFS exploratory search interface will encourage learning and navigation by showing a small initial subgraph of the most important concepts and relations. As the exploration unfolds, the user will gradually move beyond this initial subgraph.

How to construct such a knowledge subgraph? If our goal is to find a subgraph with $k$ concepts to be shown to the user (where $k$ is small), then it translates into a computational problem of selecting the best $k$-sized subgraph $H_q$ out of the original graph $G_q$. Guided by the interactive design goals \textbf{DG2} and \textbf{DG3}, we propose the following selection criteria (SC):
\begin{itemize}
    \item [\textbf{SC1}] \textbf{Relevance}. Concepts and relations in this subgraph should be centered around the user's search interest, as opposed to drifting into peripheral parts of the information space. 
    \item [\textbf{SC2}] \textbf{Coverage}. Concepts and relations in this subgraph should cover diverse subregions of the information space, as exploratory search aims at breadth and learning. 
    \item [\textbf{SC3}] \textbf{Efficiency}. The subgraph selection and update procedures should be responsive enough to support a smooth search experience \cite{arapakis2014impact}.
\end{itemize}

We note that these selection criteria are related to search result diversification~\cite{rodrygo2015search}, where the goal is to provide result documents to cover different aspects of an ambiguous query. In the learning stage of an exploratory search, the user's goal is also ambiguous. The crucial difference is that here our goal is to select or rank knowledge graph concepts instead of result documents.

Even after being pruned to a manageable size, a knowledge graph can still be too abstract to make immediate sense in a user's eyes.  To further support user interpretation of the extracted knowledge subgraph $H_q$, we augment it with the following data elements to be used in the frontend interface.

\emph{Concept Provenance}. For each concept $c \in H_q$, we need a small representative set of context windows (e.g. sentences) in result documents $D_q$ which mention the concept $c$ and the query $q$. These context windows explain why a concept is relevant in the search context.

\emph{Graph Partitioning}. The set of concepts in $H_q$ should be partitioned such that concepts in the same partition are densely connected by edges in $H_q$, and concepts between partitions are loosely connected by edges in $H_q$. For semantically close concepts in the same partition, we present them visually close to each other and use the same color coding in the interface. This can help with the chunking process in learning~\cite{miller1956magical} and nudge users to think about connections between nearby concepts according to the proximity principle in Gestalt Principles~\cite{todorovic2008gestalt}. In other words, visualizing these partitions is an implicit way of surfacing relations between concepts (\textbf{DG2}).


\subsection{Data Model and Interactive System Implementation}

In this section, we describe the interactive system for delivering the experience that GRAFS aims to achieve and the computational implementation of the underlying data model. Here our description follows the data flow: we first describe the implementation of the data model, followed by the implementation of the interactive system that presents the data model to users.

\subsubsection{Data Model Implementation}

We implemented the described data model on top of a custom-built search engine for 32.6 million PubMed abstracts, which were downloaded from the National Library of Medicine website. We use an efficient maximal pattern matching-based   algorithm~\cite{dai2008efficient,jonquet2009open} to annotate clinical concepts mentioned within each abstract by looking up terms in the SNOMED-CT vocabulary. The initial knowledge subgraph is built specifically for each user-issued search query to ensure \textit{relevance} (\textbf{SC1}). Given a search query $q$, we first extract the original query-specific knowledge graph $G_q = (C_q, R_q)$ as follows. All concepts mentioned in the set of retrieved documents by the search query $D_q$ form the extracted concept set $C_q$. We take a generic view of concept relations and  assume that a pair of concepts in $C_q$ are  related if they co-occur in any retrieved document. This generates a set of concept relations $R_q$.  We do not further consider fine-grained types of relations because, in the context of a specific query, concepts can be related in novel and nuanced ways that are not documented in an ontology (e.g., SNOMED-CT), and accurately extracting such relations from text is a challenging natural language understanding task~\cite{luo2022biored}. 

In principle, one can apply existing methods for ranking and selecting knowledge graph concepts for search tasks~\cite{dalton2013constructing,schuhmacher2015ranking,dietz2019ent}. However, these methods are mainly optimized for batch evaluation settings and their computational cost can be too high to be run at an interactive rate. For example, shortest-path algorithms and random walks on the query-specific knowledge graph $G_q$ have superlinear (e.g., quadratic) time complexity in the number of concepts in $G_q$. We adopt an efficient implementation instead.


We construct a subgraph $H_q \in G_q$ that contains a user-specified number (e.g., 20) of concepts in $C_q$. We incorporate the \emph{relevance} (\textbf{SC1}) and \emph{coverage} (\textbf{SC2}) criteria into an efficient subset selection algorithm (\textbf{SC3}). The algorithm is inspired by the maximal marginal relevance~\cite{carbonell1998use}. The basic idea is to sequentially select items that simultaneously have high relevance to the query and few relationship links between each other.


Formally, let $q$ be the search query, $C_i \subset C_q$ be the current set of selected concepts with size $i$ (initially $i=0$, $C_i = \varnothing$), and $C_q\setminus C_i$ be the current set of unselected concepts. We select the next concept  $c_{i+1}$ as follows:
\begin{equation}
c_{i+1} \leftarrow \argmax_{c \in C_q \setminus C_i} \left[\lambda \cdot r(c, q) - (1-\lambda) \cdot \max_{c_j \in C_i} s(c, c_j) \right] \ . \label{equ:mmr-concept}
\end{equation}
Here, $c$ is a candidate concept in the current set of unselected concepts. $r(c,q) = |\{d|c \in d, d \in D_q\}|$ measures the \textit{relevance} (\textbf{SC1}) of $c$ by the number of retrieved documents containing $c$. $s(c, c_j) = |\{d|c \in d, c_j \in d, d \in D_q\}|$ measures the relationship strength  between $c$ and a selected concept $c_j \in C_i$ by the number of documents where $c$ and $c_j$ co-occur. Intuitively, the term ``$-\max_{c_j \in C_i} s(c, c_j)$'' promotes \textit{coverage} (\textbf{SC2}) of the next concept $c$ by forcing it to be semantically far from the current selected set $C_i$. $0 < \lambda < 1$ defines the relative importance of relevance and coverage criteria. We set $\lambda$ to a static value ($\lambda = 0.5$) during our study to put equal importance on relevance and coverage criteria. $k$ concepts are added into $H_q$ incrementally following Equation \eqref{equ:mmr-concept}.
The algorithm has a time complexity of $O(k^2|C_q|)$. The algorithm runs \textit{efficiently} (\textbf{SC3}) since $k$, the number of concepts in the subgraph $H_q$, is usually small (e.g., 20), and the algorithm has linear (instead of quadratic) complexity in the number of concepts in $G_q$, which is often very large (e.g., 5000).

We noticed that some selected concepts are frequent but not informative. For instance, the concept ``COVID-19'' appears in more than half of the documents retrieved by a COVID-19 related query, and therefore has a large relevance term $r(c,q)$. However, such a concept contains little new information about the topic. We, therefore, omitted concepts with $r(c,q) > |D_q|/2$, treating them as equivalent to ``stop words''  (frequent functional words)  in the context of the current search results $D_q$.

\textbf{Concept Provenance.} For each selected concept $c$ in $H_q$, we extracted three representative sentences containing both concept $c$ and query $q$ to show how $c$ is used in the context of search results. The selection algorithm is also inspired by maximal marginal relevance. First, all sentences containing $c$ were extracted from $D_q$ to form the candidate sentence set $S_q$. Let $S_i \in S_q$ be the current set of selected sentences with size $i$ (initially $i=0, S_q = \varnothing$). The algorithm selects the next sentence $s_{i+1}$ as follows:
\begin{equation}
s_{i+1} \leftarrow  \argmax_{s \in S_q \setminus S_i} \left[\lambda \cdot v(s, q) - (1-\lambda) \cdot \max_{s_j \in S_i} v(s, s_j) \right] \ .
\label{mmr-sentence}
\end{equation}
Here, $v(s,q)$  measures the relevance of sentence $s$ with respect to query $q$, while $v(s, s_j)$  measures the similarity between the candidate sentence $s$ and a selected sentence $s_k$. The function $v(\cdot, \cdot)$ represents queries and sentences as bag-of-words vectors with TFIDF weights and computes cosine similarity between two vectors. We empirically set $\lambda = 0.5$ to balance the relevance and diversity of the set of selected sentences.

\textbf{Graph Partitioning.} This step generates groups of concepts in $H_q$ based on their relations. Instead of choosing a fixed number of partitions (as in algorithms like $k$-means or $k$-medoids), we performed agglomerative hierarchical clustering over the concepts in $H_q$. First, each concept starts in its own cluster. At each subsequent step, the two closest clusters are merged. The distance between two clusters is the distance between the farthest concepts (complete-linkage clustering). The distance between two concepts is calculated as the symmetric difference between the two sets of documents that contain each concept: $d(c_i, c_j) = |D_i \cup D_j| - |D_i \cap D_j|$, where $D_i = \{d|c_i\in d, d \in D_q\}$. $d(c_i, c_j) = 0$ when $D_i = D_j$, i.e., when $c_i$ and $c_j$ always co-occur in retrieved documents and are therefore very close semantically. Therefore, even if two concepts $c_i, c_j$ frequently co-occur ($|D_i \cap D_j|$ is large), their distance is still far if their union is much larger than their intersection. This avoids the problem where a concept is viewed to be close to many other concepts simply because it appears in many documents (e.g., concepts in the query that are prevalent in search results).

\begin{figure}[t]
\centering  
\begin{subfigure}{.45\textwidth}
    \centering
    \includegraphics[width=1\textwidth]{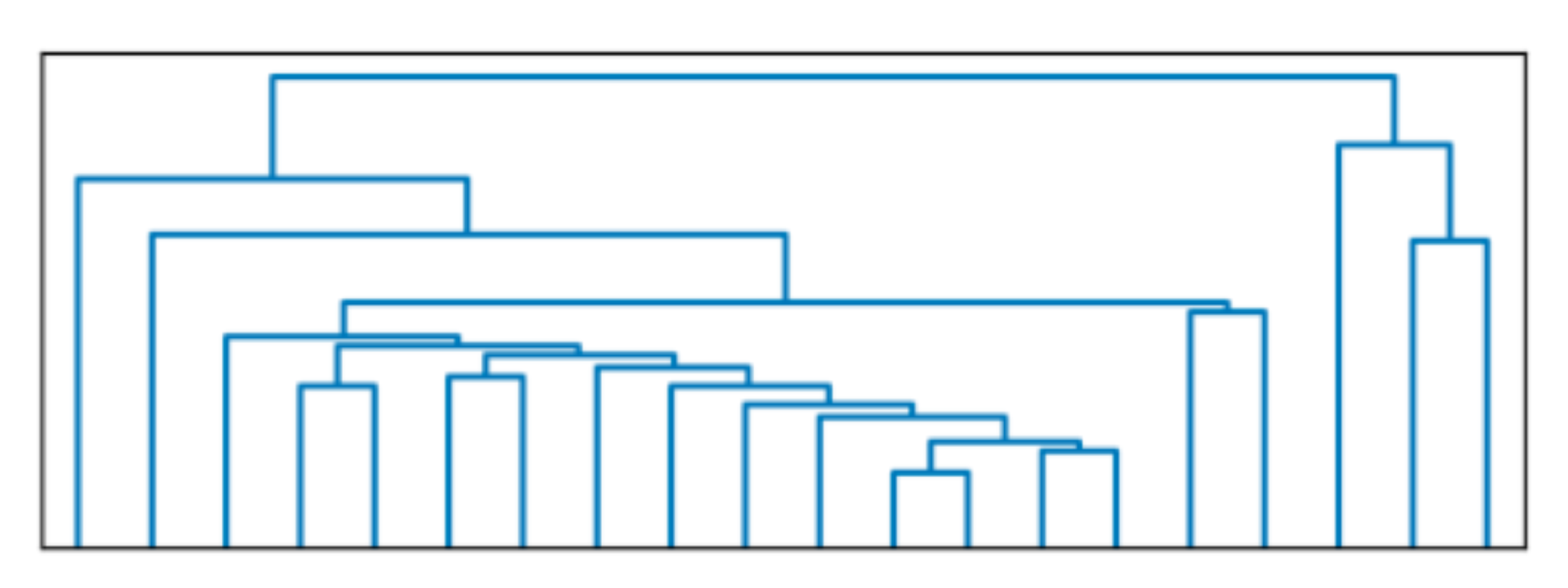}
    \caption{The dendrogram generated by hierarchical clustering}\label{fig:dendro_a}
\end{subfigure}
    \hfill
\begin{subfigure}{.46\textwidth}
    \centering
    \includegraphics[width=1\textwidth]{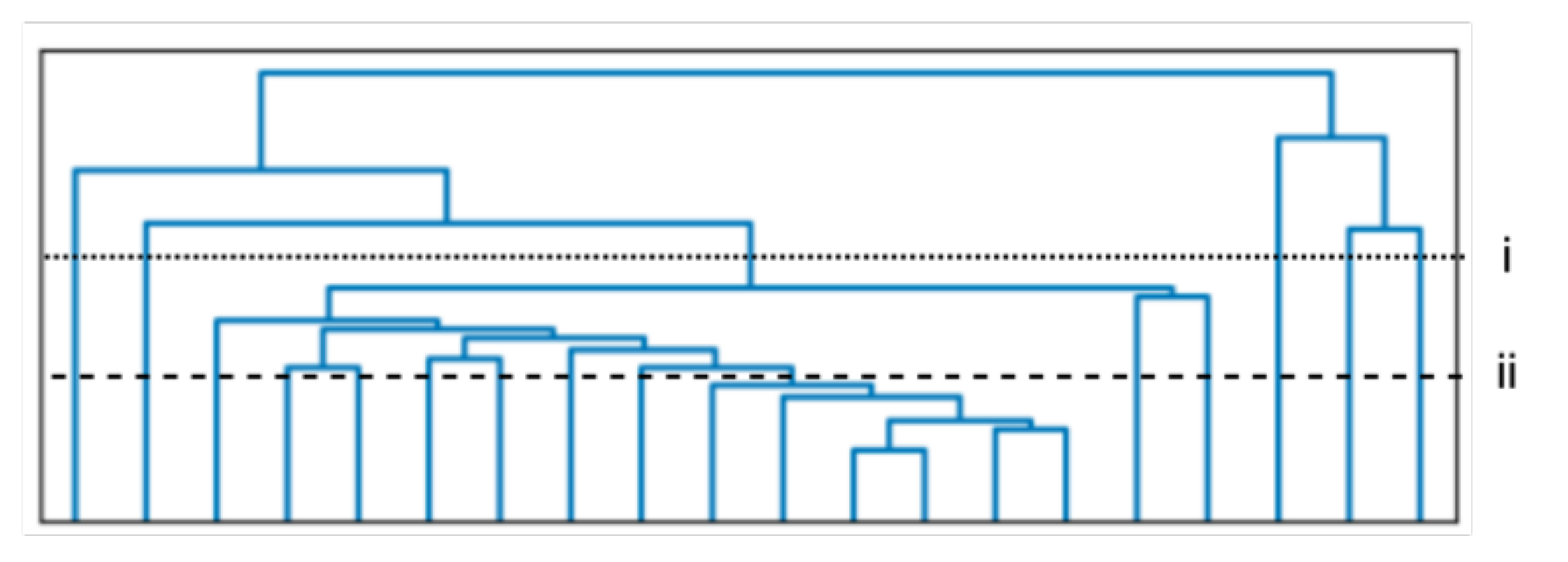}
    \caption{Cutting the hierarchy at a fixed distance}\label{fig:dendro_b}
\end{subfigure}
    \hfill
\begin{subfigure}{.45\textwidth}
    \centering
    \includegraphics[width=1\textwidth]{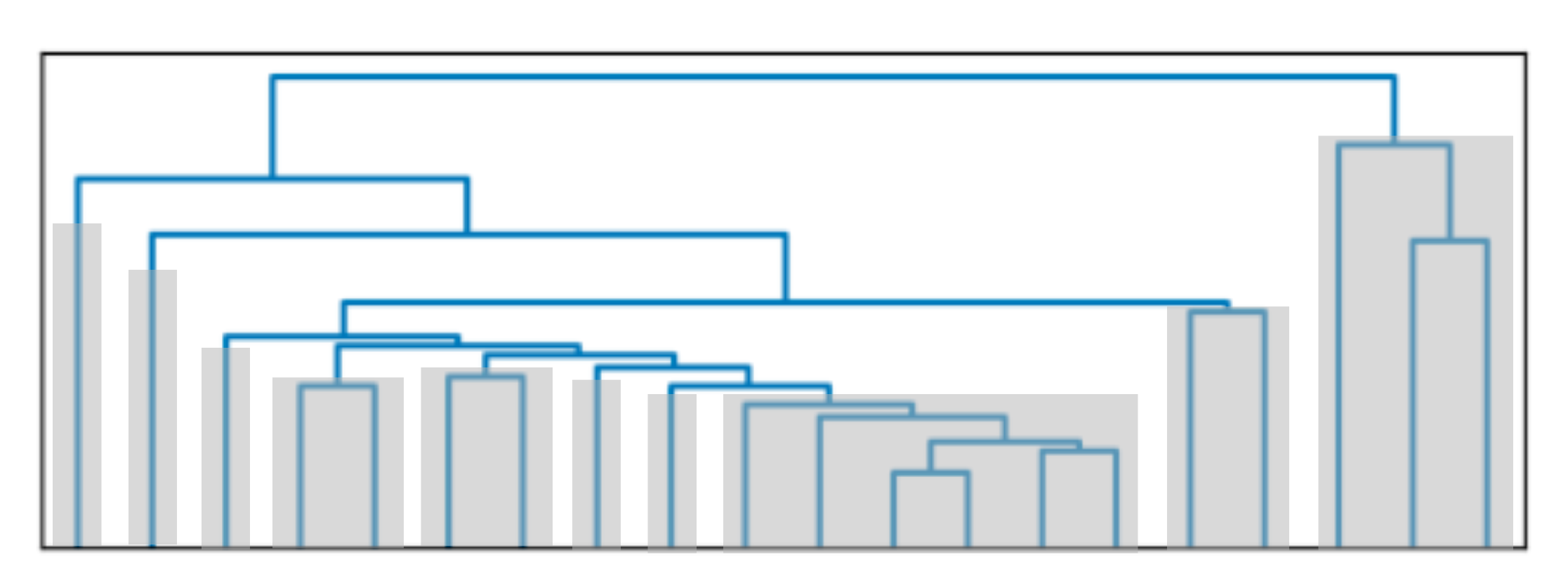}
    \caption{Our method of cutting the hierarchy}\label{fig:dendro_c}
\end{subfigure}
    \hfill
\begin{subfigure}{.45\textwidth}
    \centering
    \includegraphics[width=1\textwidth]{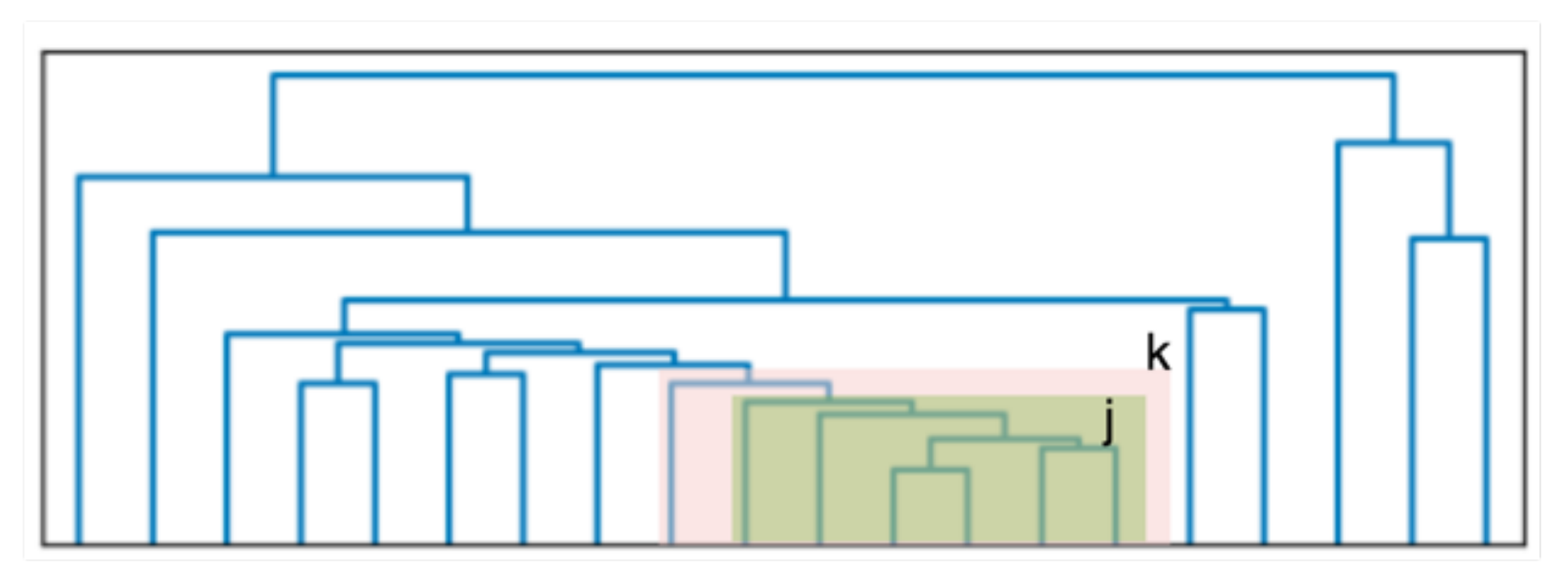}
    \caption{Identifying the largest clusters that is smaller than one third of
the total number of nodes}\label{fig:dendro_d}
\end{subfigure}
\caption{(a) The hierarchical clustering algorithm produced a tree structure. (b) Cutting the hierarchy at a fixed distance level can easily generate one very large cluster (i) or numerous small clusters (ii). (c) We cut the tree such that each sub-tree is as large as possible but has no more than one third of all leaf nodes. (d) Cluster j is a valid cluster as including one more node will result in a cluster (Cluster k) with seven nodes, exceeding one third of the total number of leaf nodes (20 nodes in total).}
\label{fig:graph_partitioning}
\end{figure}

The hierarchical clustering algorithm produced a tree structure where each leaf corresponds to a concept. To generate partitions, we cut the tree such that each  sub-tree is as large as possible but has no more than one-third of all concepts (leaf nodes) in the original tree. This method is able to produce partitions with relatively balanced sizes regardless of the original tree structure. Each partition contains semantically related concepts. This approach is different from the commonly used method of cutting the tree at a fixed level, which may produce one very large partition of loosely related concepts or numerous small partitions if the original tree structure is highly skewed. Figure \ref{fig:graph_partitioning} illustrates a concrete example of our partitioning method. 

\subsubsection{Interactive System Implementation}
We implement GRAFS as an interactive web application. Its frontend elements leverage the JavaScript library D3.js. Its backend document indexing and search functions are based on Apache Solr.
At the beginning of a search,  a user can issue her search query of interest in the same manner as in a typical search engine. Simple keyword queries or complex Boolean queries are supported.  The user can also specify the number of documents ($|D_q|$) to retrieve. We set the default number of $|D_q|$ to be 1000, since we aimed to assist users in the exploration of a large and complex information space. 
After retrieving a set of documents, the system backend will compute the data model (knowledge subgraph, concept provenance, and graph partitioning) on the fly, which will serve as the input to the frontend interface. The user can specify the desired number of concepts in the knowledge subgraph, which by default is set to 20. The default number 20 was determined empirically during prototype development to provide an informative set of concepts while also not overwhelming the user with too much information.

\begin{landscape}
\begin{figure}[t]
\includegraphics[width=1\columnwidth]{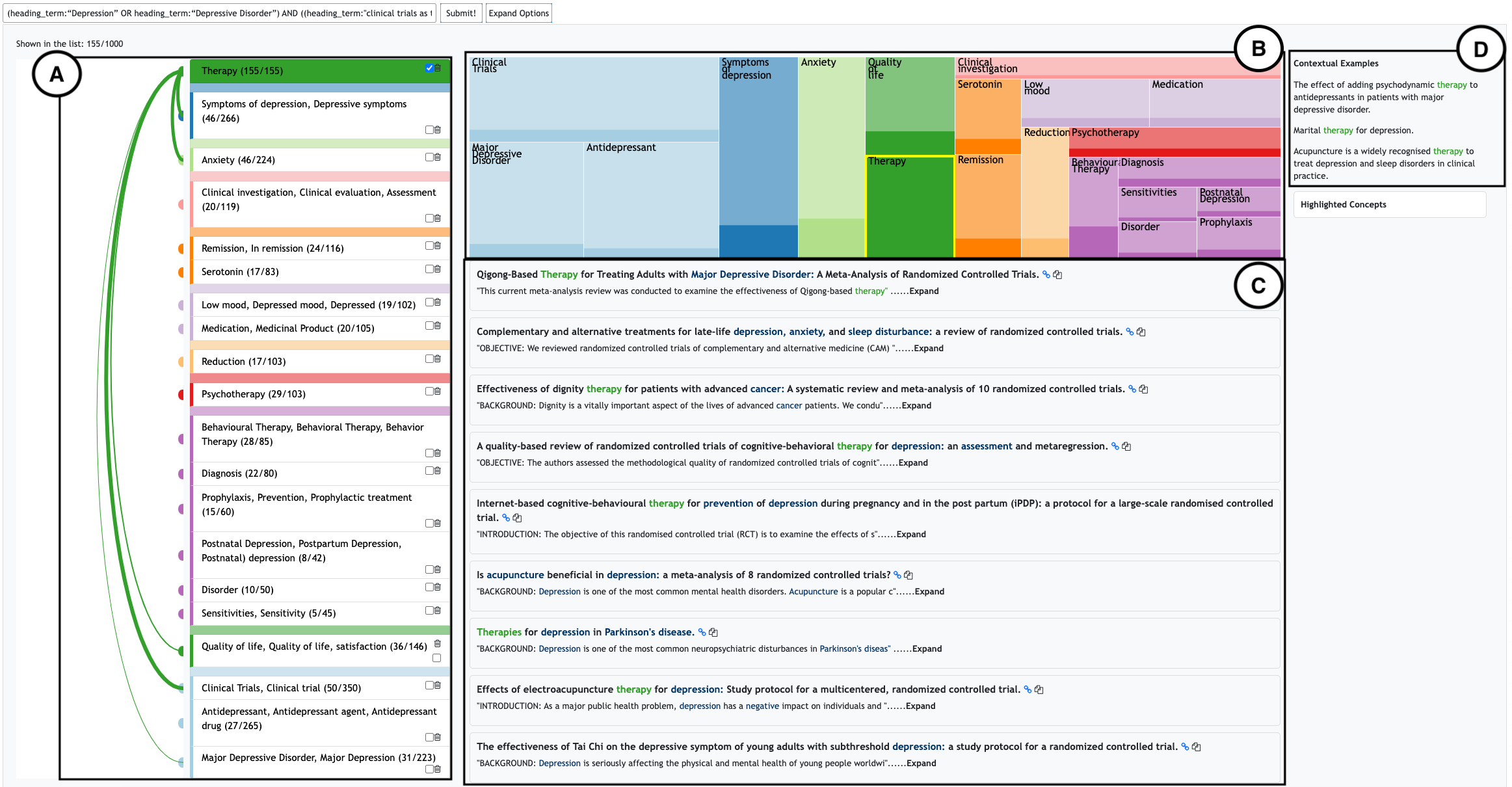}
\Description[Initial view]{Initial view}
\caption{GRAFS for the topic ``treatment for depression'', including (A) the \textbf{graphical facet list} containing key concepts and their relations represented by arcs, (B) the \textbf{graphical facet treemap} showing the hierarchical structure of concepts based on the graph partitioning results, (C) the \textbf{document list} showing filtered articles ranked by their relevance to the search query, (D) the \textbf{concept provenance} that lists three examples of the selected concept. }
\label{fig:action_view}
\end{figure}
\end{landscape}

\textbf{Interface Overview}.
A demonstration video showing all features of the GRAFS interface is \href{https://drive.google.com/file/d/1T36r-XxDQ8Y31AD1JFe6yUqtla_qvtR0/view?usp=sharing}{available online}\footnote{\url{https://tinyurl.com/m3nabas3}
}.
Figure \ref{fig:action_view} shows the interface rendered for the search topic ``treatment for depression'' with 20 selected concepts. The corresponding search query is shown in the search box at the top of the interface, where the user can construct or edit search queries.

To support exploratory search, GRAFS user interface (UI) follows the structure of a faceted search interface with a \textit{facet list} (Figure \ref{fig:action_view}(A)) on the left and a \textit{document list} (Figure \ref{fig:action_view}(C)) in the middle (\textbf{DG1}). In Figure \ref{fig:action_view}(A), each facet is a concept/node of the initial knowledge subgraph, which allows users to learn major concepts and filter articles by those concepts. The \textit{facet treemap} (Figure \ref{fig:action_view}(B)) shows a space-filling treemap~\cite{shneiderman1992tree} of  concepts, which visualizes the prevalence and relationship with each other. Given a selected concept (``Therapy'' in Figure \ref{fig:action_view}), the concept provenance sentences are listed on the right (Figure \ref{fig:action_view}(D)), providing a succinct summary of the concept.

\begin{figure}[h]
\centering  
\begin{subfigure}{.28\textwidth}
    \centering
    \includegraphics[width=1\textwidth]{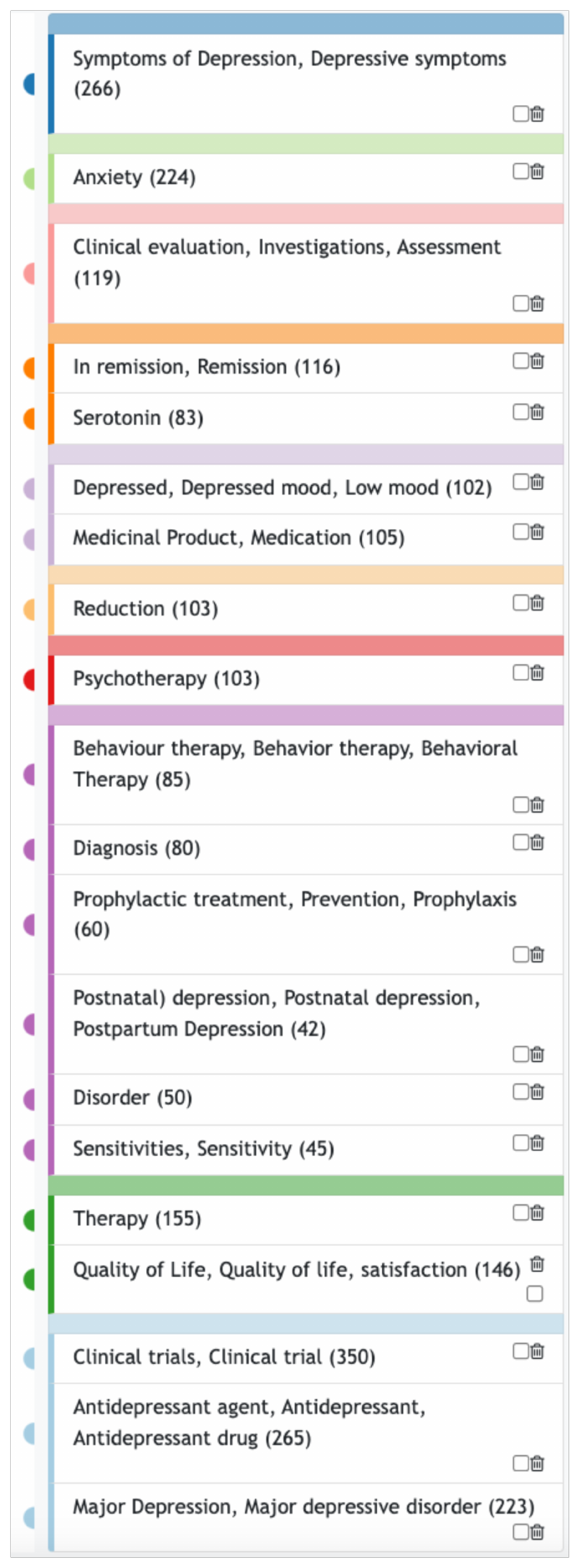}
    \caption{Initial view}\label{fig:dendro_a}
\end{subfigure}
    \hfill
\begin{subfigure}{.35\textwidth}
    \centering
    \includegraphics[width=1\textwidth]{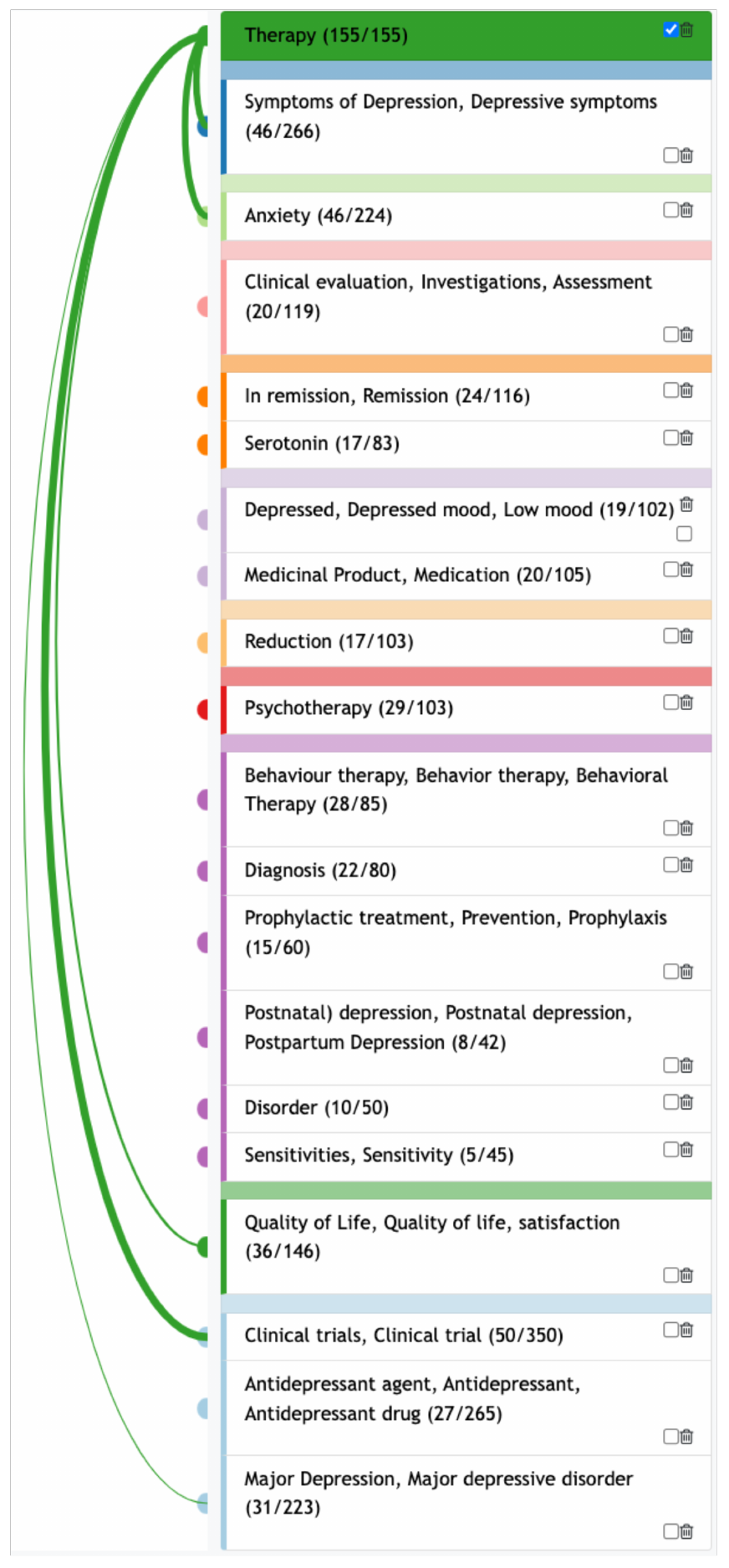}
    \caption{Select "Therapy"}\label{fig:dendro_b}
\end{subfigure}
    \hfill
\begin{subfigure}{.35\textwidth}
    \centering
    \includegraphics[width=1\textwidth]{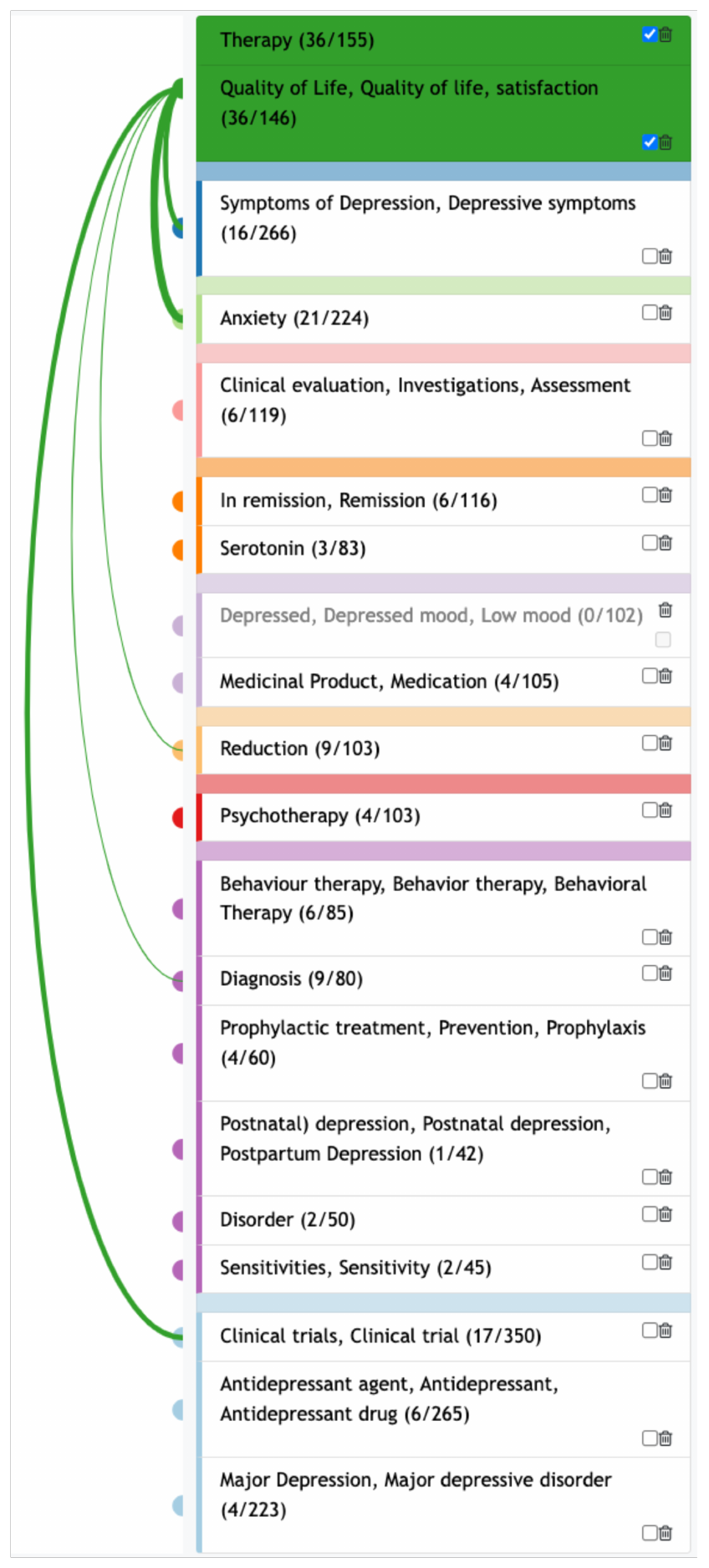}
    \caption{Select "Therapy" and "Quality of Life"}\label{fig:dendro_c}
\end{subfigure}
\caption{Facet list transitions. (a) The user starts with an initial view. (b) After the concept ``Therapy'' is selected, it is moved to the top of the facet list. Arcs are drawn to connect the selected concept to the five most related concepts. The arcs have the same color as the selected concept. (c) A second concept ``Quality of Life'' is selected, and it is also moved to the top of the facet list. Relations between the intersection of the selected two concepts (at the top) and the other concepts are again shown by arcs.}
\label{fig:concept_list}
\end{figure}

\textbf{1D Graphical Facet: Facet List}.
Figure \ref{fig:concept_list}(a) shows an initial view of the facet list.
The one-dimensional facet list contains the concepts/nodes in the knowledge subgraph. Based on the result of graph partitioning, the order of concepts follows the  order of leaves given by the cluster hierarchy and the color of each concept indicates its group identity. 

Users can select a concept to filter the retrieved document set so that only documents containing the selected concept will be listed in the document list. For instance, the user might be interested in ``Therapy.'' Upon selection, the concept (``Therapy'' in Figure \ref{fig:action_view} and \ref{fig:concept_list}(a)) will appear at the top of the list. The rest of the concepts maintain the same order as the initial facet list to maintain the concept grouping information. 

The visualization component in the facet list was also designed to facilitate users' exploration of the knowledge subgraph and construction of their own mental model. To achieve this, arcs are drawn to connect the selected concept to the top five most related concepts in the knowledge subgraph. 
The choice of ``top five'' was determined empirically during prototype development as a compromise between the desire to provide information and the need to avoid excessive visual complexity that could reduce usability. Though the users in our evaluation study appeared satisfied with the choice of five arcs (no negative feedback was provided regarding this design choice), other thresholds or heuristic-based approaches are possible alternatives.  
The top arc connections allow users to inspect a small portion of the per-query knowledge subgraph given the current search focus and get a visual hint about possibly interesting concepts to explore. For instance, 
using the co-occurrence frequency to indicate relationships between concepts, the five concepts most related to ``Therapy'' are connected with arcs drawn with a thickness that is proportional to the corresponding co-occurrence frequency. For example, ``Therapy'' and its most frequently co-occurring concept ``Clinical Trials'' appear together in 50 documents. This is reflected by the thickest arc in the example which is drawn to connect these two concepts.

The interface also allows users to further narrow down the search results by applying multiple filters. When more than one concept is selected, only documents that contain all the selected concepts will be included in the interface. For instance, the user may observe that ``Quality of Life'' was clustered into the same group as ``Therapy'', but their connection is not as strong as other concepts such as ``Clinical Trials''. To answer the question, the user applied ``Quality of Life'', which allows the user to examine the documents where the two concepts co-occur. Further reading helped the user find several articles that mention ``Quality of Life'' as an important secondary outcome for measuring the effectiveness of ``Therapy''. The relation between the two selected concepts and the rest concepts are again shown by arcs, where the thickness indicates co-occurrence frequency, as shown in Figure \ref{fig:concept_list}(c). Based on those arcs, the user may hypothesize that ``Symptoms of Depression'' and ``Anxiety'' can also serve as measures of the effectiveness of ``Therapy'', and these measures may be used in ``Clinical trials''. They can test their hypothesis by further selection and reading. Therefore, these arcs in the facet list visually nudge the user to narrow down their interest by selecting the next concept.

\begin{figure}[t]
\centering  
\begin{subfigure}{0.6\textwidth}
    \centering
    \includegraphics[width=1\textwidth]{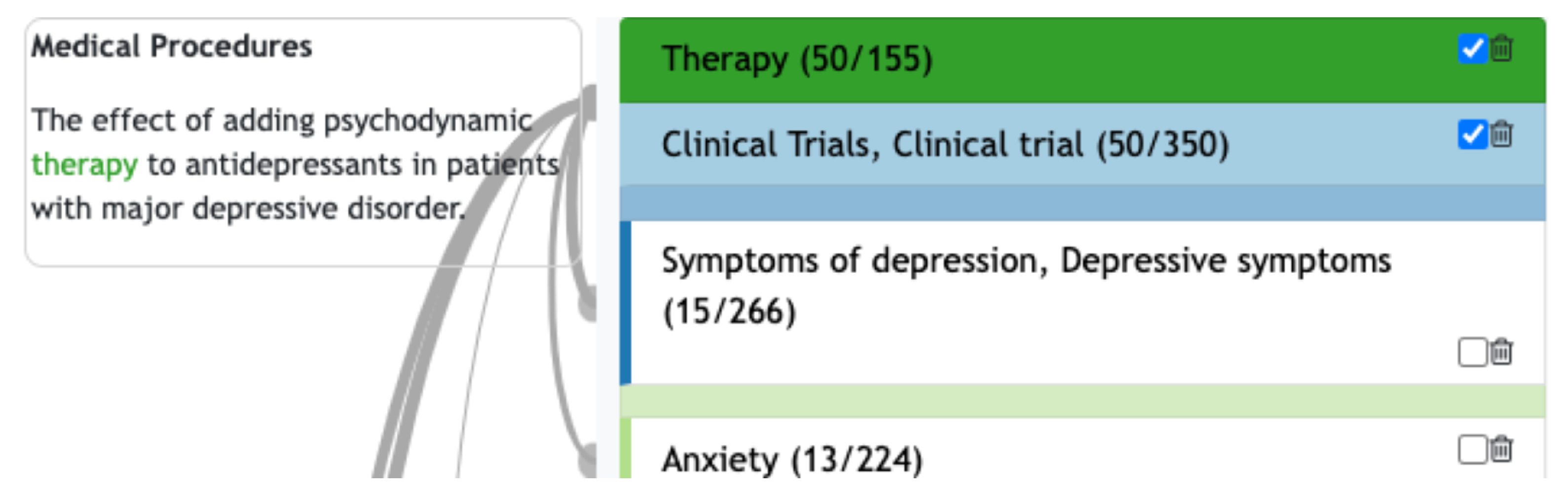}
    \caption{Example tool-tip}\label{fig:example_a}
\end{subfigure}
    \hfill
\begin{subfigure}{0.3\textwidth}
    \centering
    \includegraphics[width=1\textwidth]{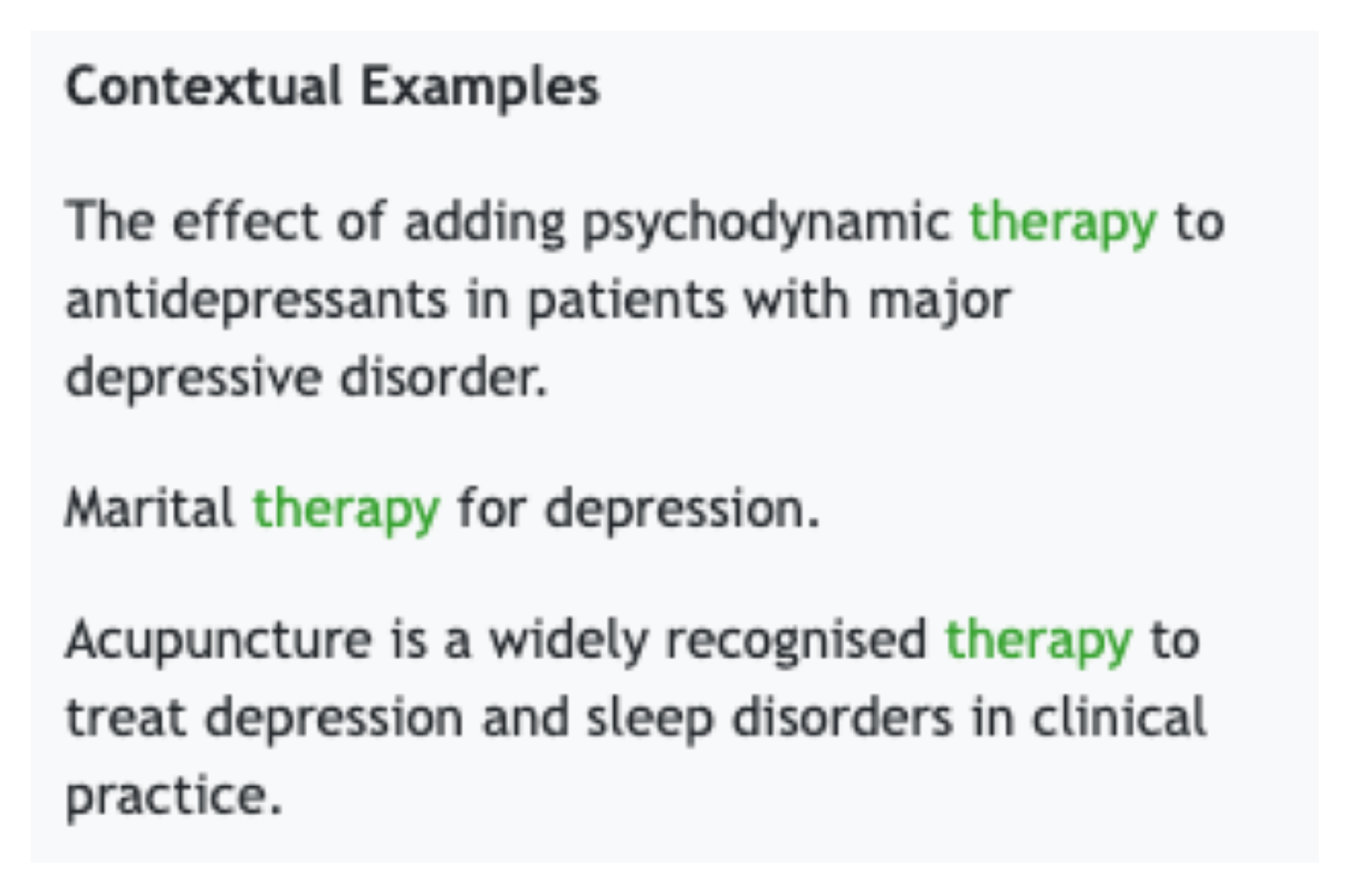}
    \caption{Contextual examples}\label{fig:example_b}
\end{subfigure}
\caption{Concept provenance. In (a), as the user hover over a concept in the facet list, the type of the concept and an example are shown in a tooltip next to the concept. In (b), when the user selects a concept, three examples are shown to provide a brief summary of how the concept is used in context.}
\label{fig:example}
\end{figure}

When the user hovers over a concept in the facet list, a tooltip with the concept's example sentence appears to the left of the facet list as shown in Figure \ref{fig:example} (a). At the same time, all the arcs that are not connected with the hovered concept turn grey to highlight only the arcs connecting to the hovered concept.
When multiple concepts from the same cluster are selected, the arcs have the same color as the cluster (Figure \ref{fig:concept_list}(c)). When multiple concepts from different clusters are selected, the arcs are grey, a color that is not used by any cluster (Figure \ref{fig:example} (a)).


\textbf{Concept Provenance}.
Concept provenance sentences are included as (1) the tooltip shown to the left of the facet list when the user hovers over a concept and (2) contextual examples listed on the right. The tooltip only includes the most related sentence extracted, aiming at providing a quick explanation of the concept as the user goes through the concept list, as shown in Figure \ref{fig:example}(a). Users can refer to all three extracted representative sentences in contextual examples, as shown in Figure \ref{fig:example}(b). 

\begin{figure}[h]
\centering  
\begin{subfigure}{1\textwidth}
    \centering
    \includegraphics[width=1\textwidth]{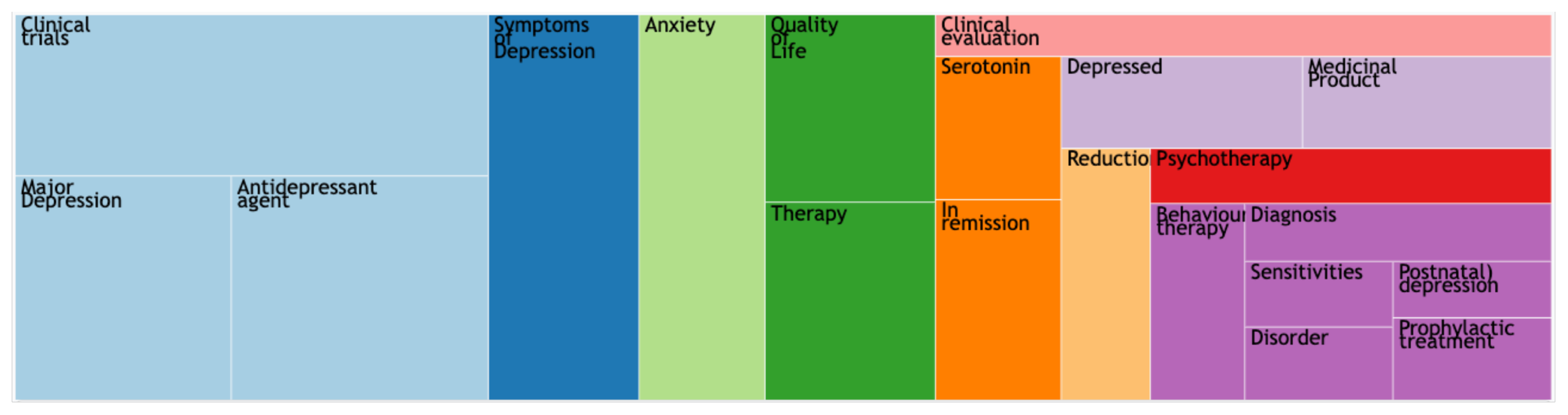}
    \caption{Initial view}\label{fig:tree_a}
\end{subfigure}
    \hfill
\begin{subfigure}{1\textwidth}
    \centering
    \includegraphics[width=1\textwidth]{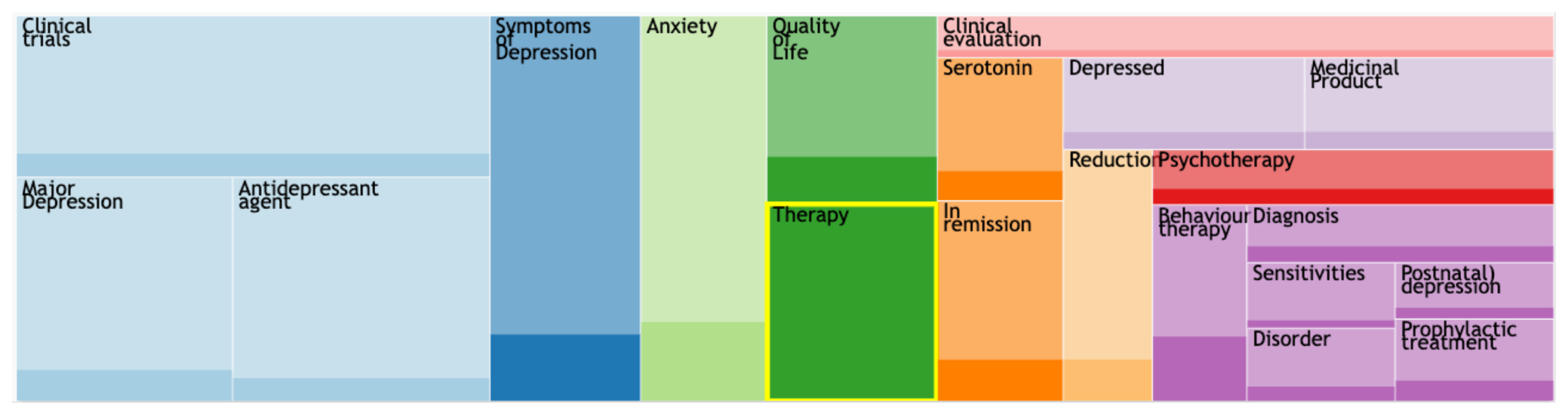}
    \caption{Select "Therapy"}\label{fig:tree_b}
\end{subfigure}
    \hfill
\begin{subfigure}{1\textwidth}
    \centering
    \includegraphics[width=1\textwidth]{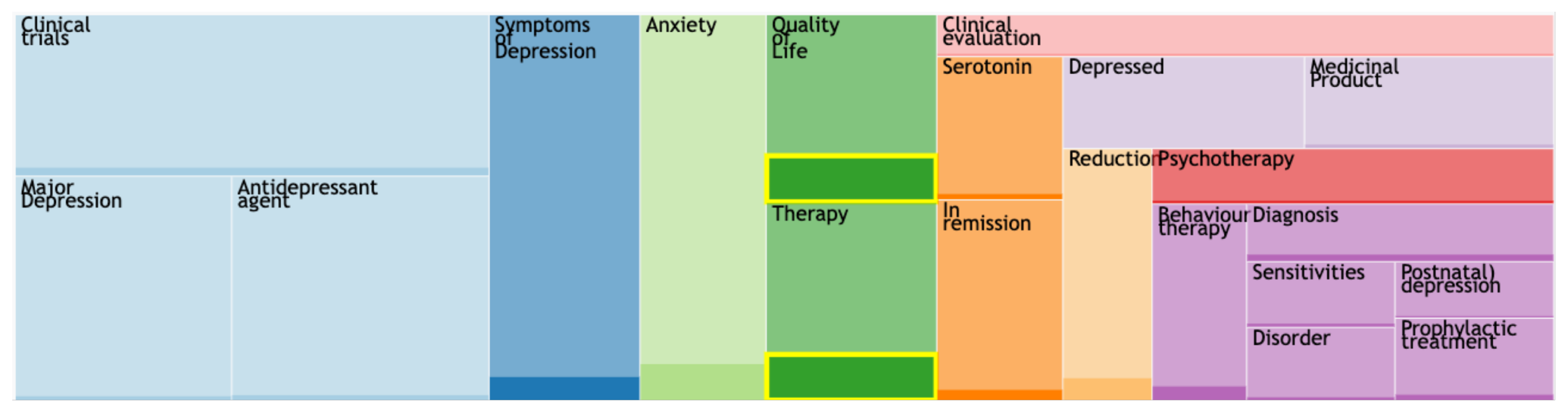}
    \caption{Select "Therapy" and "Quality of Life"}\label{fig:tree_c}
\end{subfigure}
\Description[Initial view]{Initial view}
\caption{Facet treemap transitions by doing the same sequence of actions as in Figure  \ref{fig:concept_list}. The user first selects ``Therapy'' from the initial view (a). In (b), the dark bar in each rectangle shows the overlap with the selected concepts. The user continues to select ``Quality of Life'', giving rise to (c). }
\label{fig:treemap}
\end{figure}

\textbf{2D Graphical Facet: Facet Treemap}.
To provide a spatial overview of the selected concepts, the GRAFS UI also includes the facet treemap, which directly shows the hierarchical structure given by graph partitioning. Each rectangle represents a concept. The rectangles are colored to match the concepts' group identities, and the size of the rectangle reflects the prevalence of the corresponding concept. Figure \ref{fig:treemap} illustrates the transition of the facet treemap through an example. A treemap representation was selected for two reasons. First, it naturally encodes the clustering of concepts in a way that communicates relative frequencies of occurrence. Second, it has a relatively simple visual structure even when large numbers of concepts are included. This is in contrast to alternatives such as node-link diagrams which can be difficult to interpret as the number of nodes and edges increase.

In addition to communicating concept prevalence, the facet treemap also provides an alternative way for users to explore concept relations. The position of concepts gives clues to the distance between concepts. Moreover, when the user selects concepts, a histogram is included in each concept's rectangle to show the percentage of documents containing the selected concepts. For instance, Figure \ref{fig:treemap} (b) shows the facet treemap after selecting ``Therapy'', and the darker area in ``Clinical Trials'' shows the percentage of documents mentioning ``Therapy'' among all the documents containing ``Clinical Trials.'' The additional use of color does add some visual complexity to the visualization. However, the interactive nature of this feature, which links a user's clicks to the corresponding color changes, reduces the chance of confusion. Participants in the evaluation study did not have any difficulty interpreting the visual representation of this feature.

\begin{figure}[h]
\centering  
\begin{subfigure}{1\textwidth}
    \centering
    \includegraphics[width=1\textwidth]{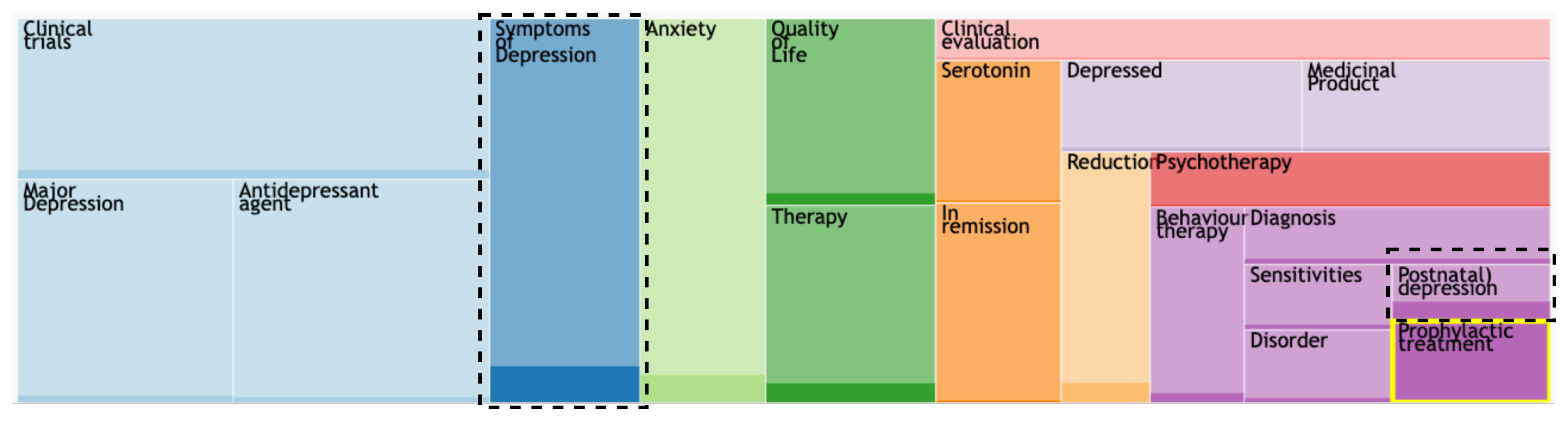}
    \caption{Select "Prophylactic treatment" in the facet treemap}\label{fig:treat_a}
\end{subfigure}
    \hfill
\begin{subfigure}{0.5\textwidth}
    \centering
    \includegraphics[width=1\textwidth]{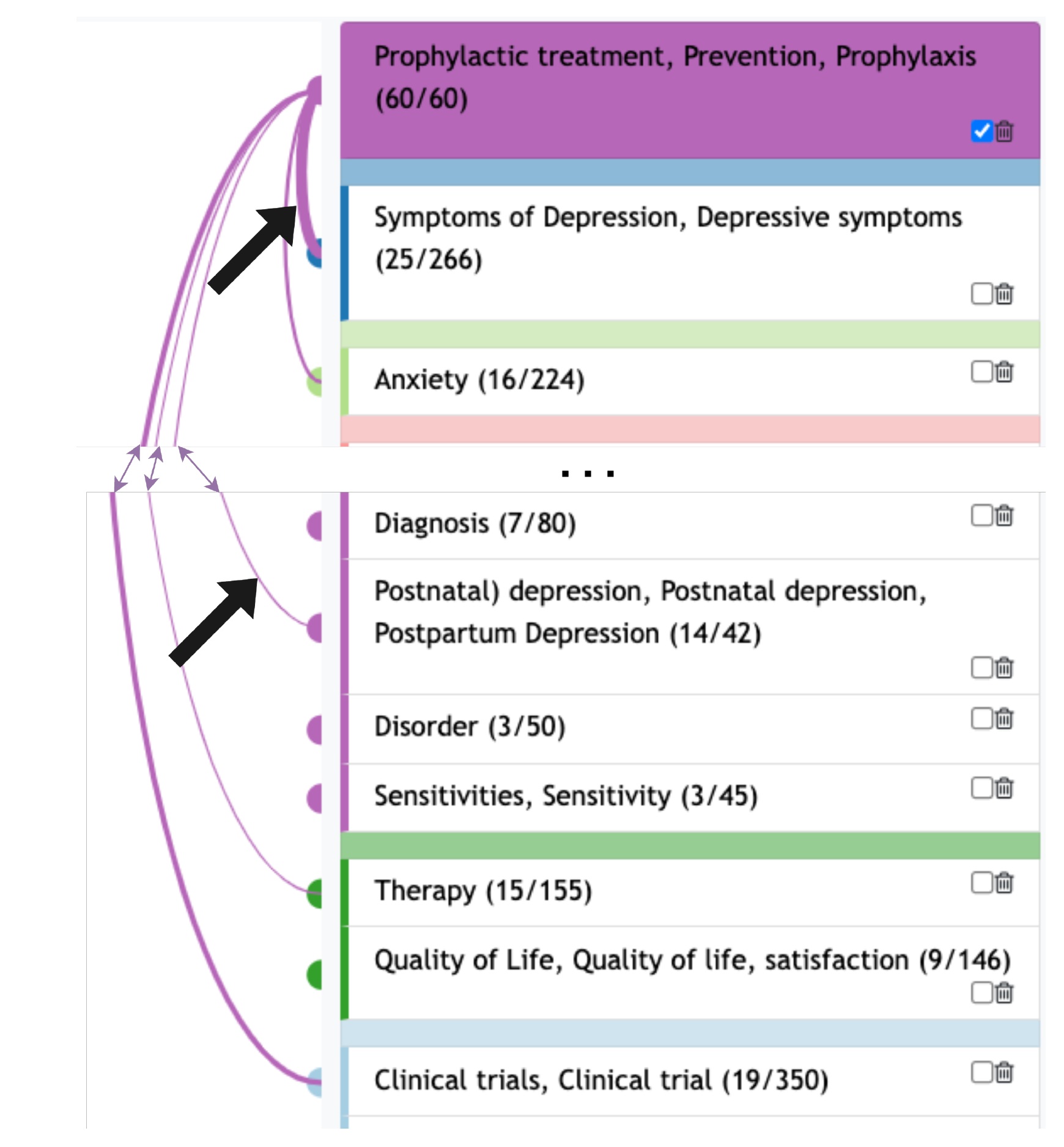}
    \caption{Select "Prophylactic treatment" in the facet list}\label{fig:treat_b}
\end{subfigure}
\Description[prevent treatment]{prevent treatment}
\caption{The facet treemap may help surface interesting relations hidden by arcs. The strong relation between ``Prophylactic treatment'' and ``Postnatal depression'' shown in (a) is not obvious in (b).}
\label{fig:prevent_treatment}
\end{figure}

Different from the arc connections indicating the co-occurrence frequency, those percentage histograms reflect concept relations relative to concept prevalence (\textbf{DG2}). This property helps surface interesting relations that might be hidden by arcs, especially for less frequent concepts. For instance, after selecting ``Prophylactic treatment'', the darker area fills a relatively large percentage of the rectangle of ``Postnatal depression'', indicating that  ``Postnatal depression'' might be closely related to ``Prophylactic treatment'' (Figure \ref{fig:prevent_treatment} (a)). And this relation seems to be stronger compared to ``Symptoms of Depression'', since the percentage covered by the darker area in the ``Symptoms of Depression'' rectangle appears to be smaller than in the ``Postnatal depression'' rectangle. This relation can be easily ignored if the user only looks at the arcs, where the connection of ``Prophylactic treatment'' with ``Symptoms of Depression'' seems to be much stronger than with ``Postnatal depression'' based on the absolute number of co-occurrences (Figure \ref{fig:prevent_treatment} (b)). 

As the facet list and the facet treemap contains the same set of concepts, we use brushing techniques to synchronize interactions on either of them. Users can select a concept by directly clicking on it through the facet treemap, and the facet list will be re-rendered to incorporate the selection. Hovering over a concept on the facet treemap also has the same effects as hovering over the facet list.

\begin{figure}[h]
\centering  
\begin{subfigure}{0.45\textwidth}
    \centering
    \includegraphics[width=1\textwidth]{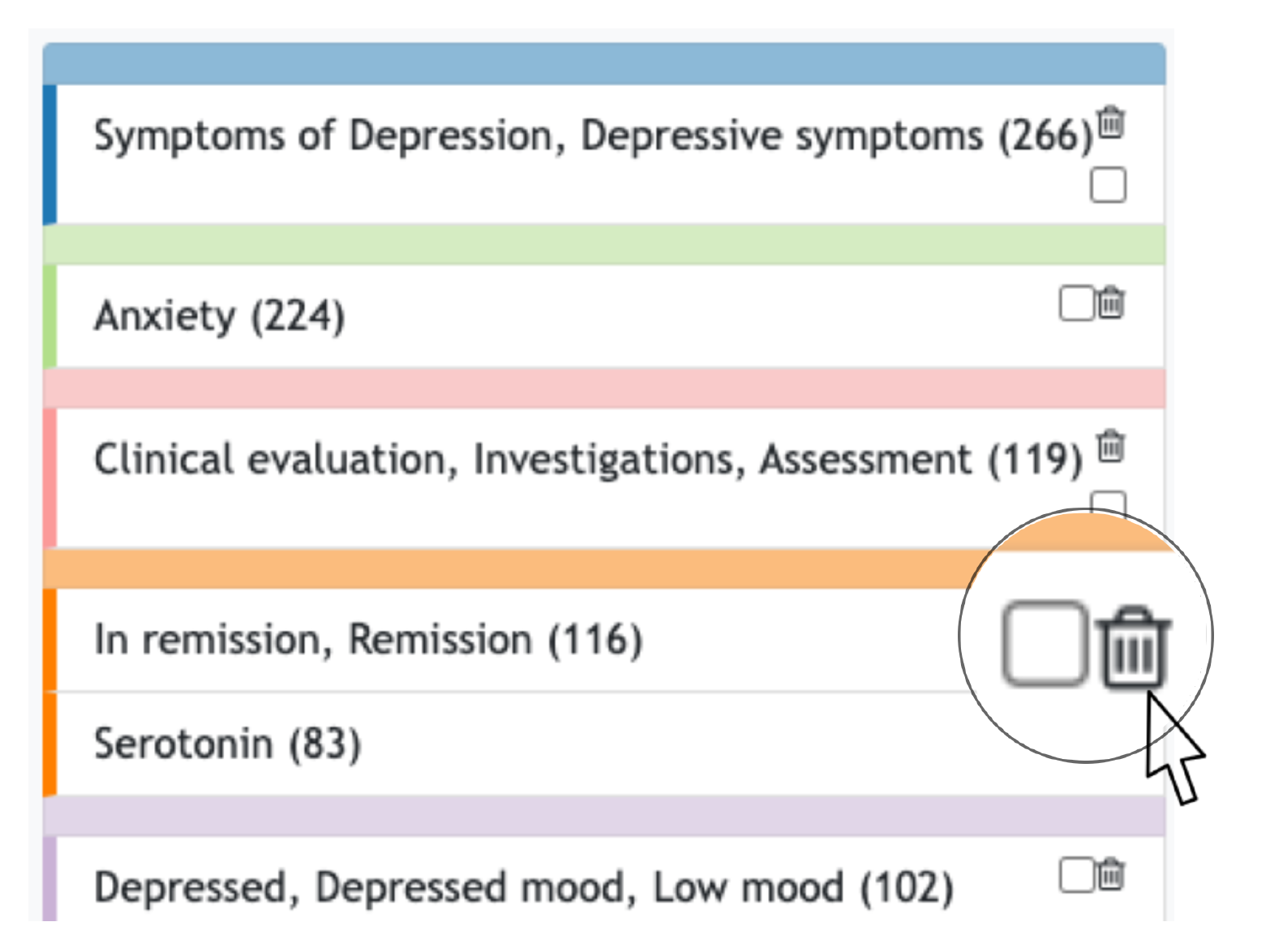}
    \caption{Delete a concept from the facet list }\label{fig:edit_a}
\end{subfigure}
    \hfill
\begin{subfigure}{0.45\textwidth}
    \centering
    \includegraphics[width=1\textwidth]{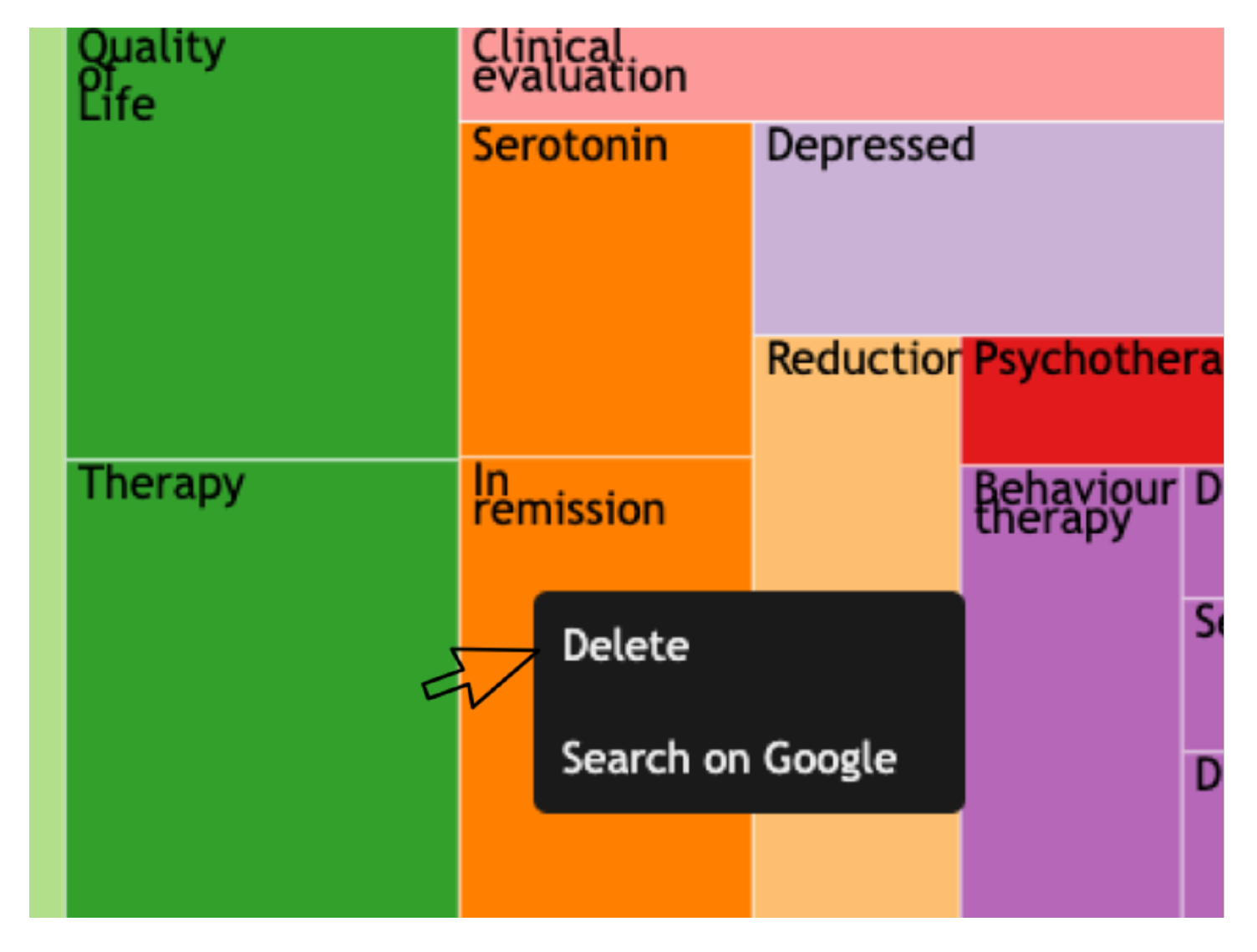}
    \caption{Delete a concept from the facet treemap}\label{fig:edit_b}
\end{subfigure}
\begin{subfigure}{0.45\textwidth}
    \centering
    \includegraphics[width=1\textwidth]{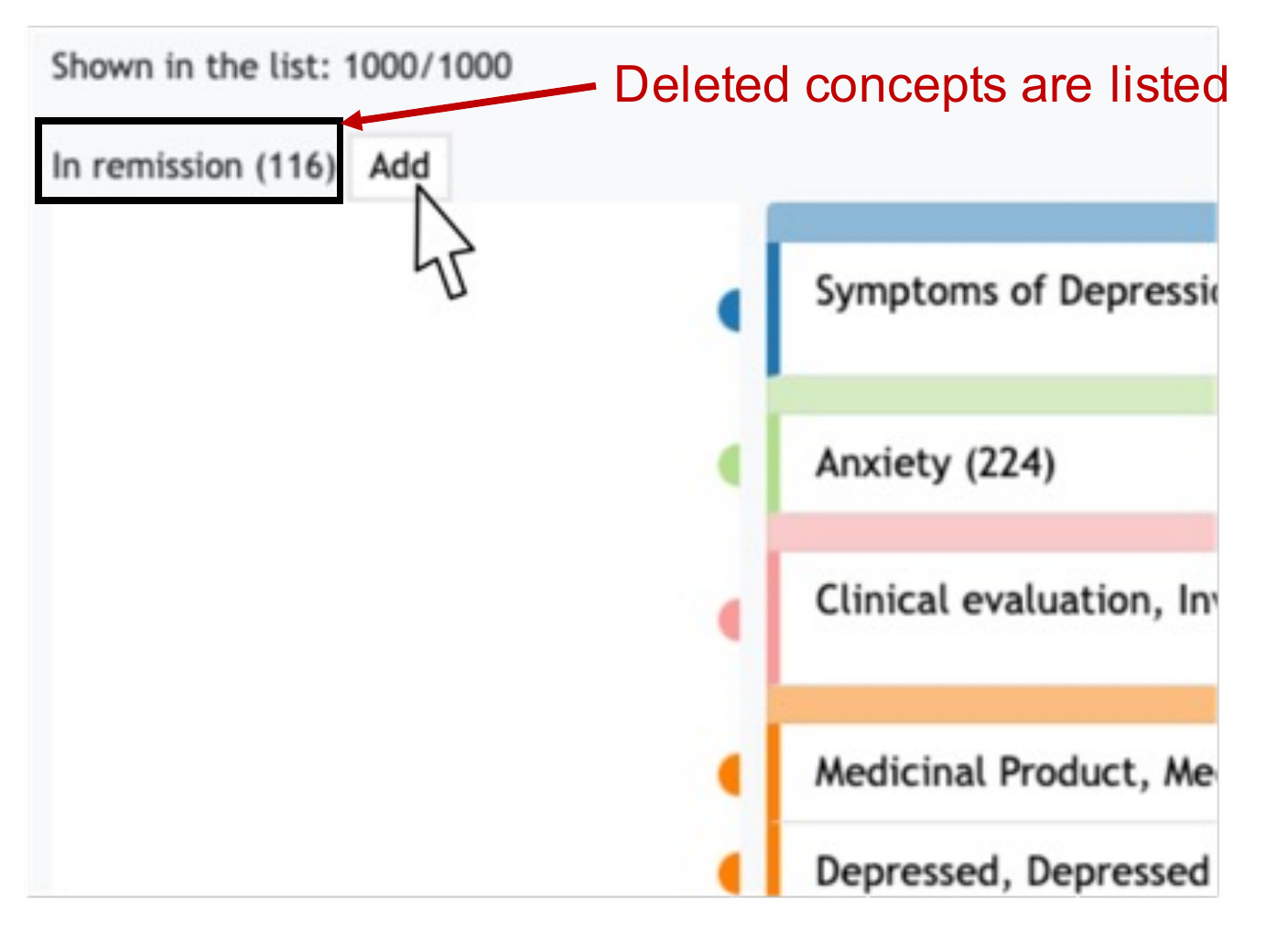}
    \caption{Add a deleted concept back to facet list}\label{fig:edit_c}
\end{subfigure}
    \hfill
\begin{subfigure}{0.45\textwidth}
    \centering
    \includegraphics[width=1\textwidth]{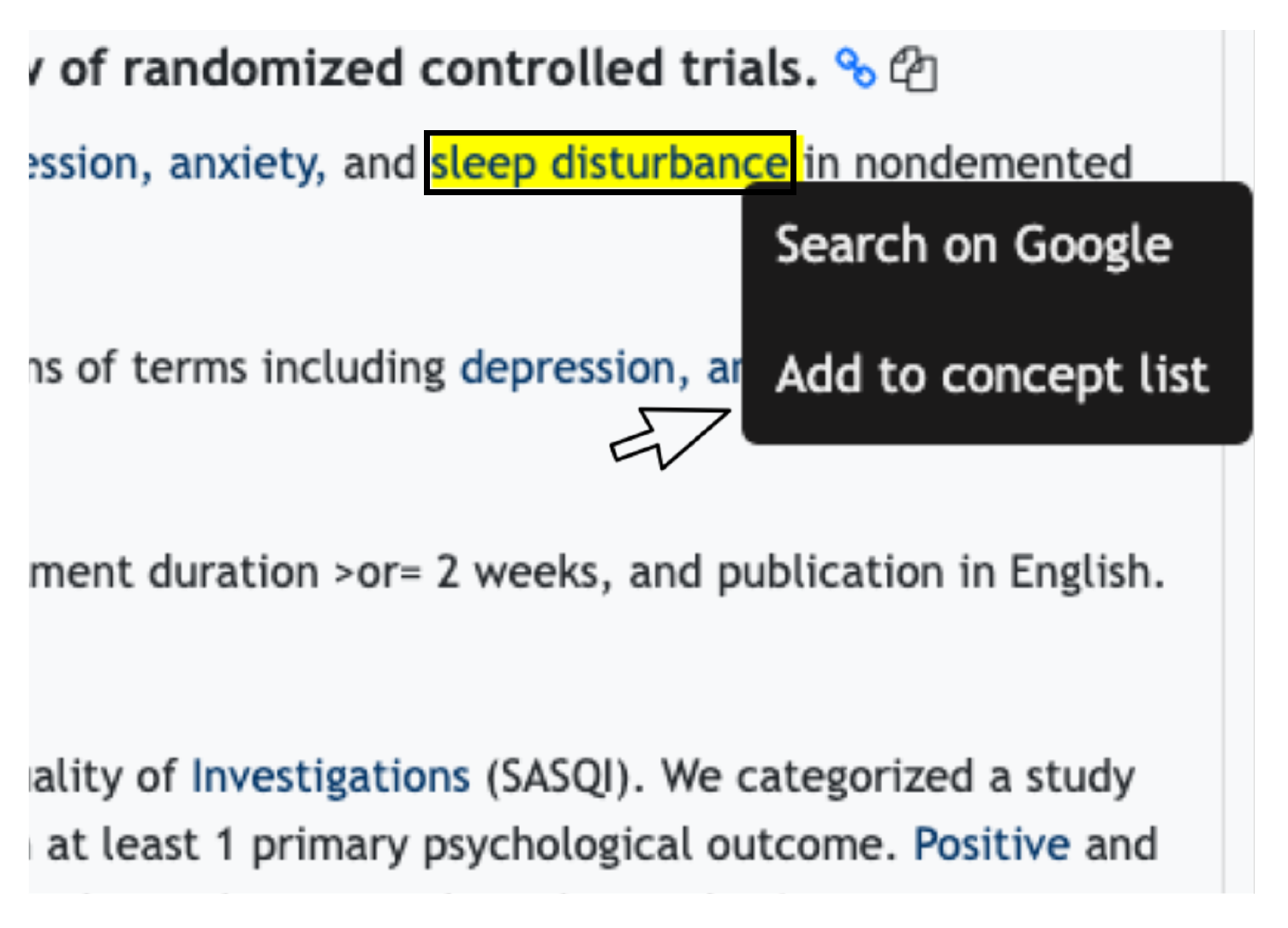}
    \caption{Add a concept from document}\label{fig:edit_d}
\end{subfigure}
\Description[edit]{edit}
\caption{Deletion can either be done through the facet list (a) or the facet treemap (b). The deleted concepts will be listed separately in the interface at the top of the facet list, and users may add them back by clicking on the "Add" button as in (c). (d) Users can add recognized SNOMED concepts to the facet list from documents.}
\label{fig:edit_list}
\end{figure}

\textbf{Editable List}.
Users may want to make changes to the facet list based on their domain knowledge or new discoveries during their exploration. As shown in Figure \ref{fig:edit_list}, users can add new concepts or remove existing concepts from the list. Deletion can either be done through the facet list or the facet treemap. As shown in Figure \ref{fig:edit_list} (c), the deleted concepts will be listed separately in the interface at the top of the facet list in case the user wants to add the concept back. Users may encounter interesting concepts in articles during exploration and want to gain a deeper understanding of them. To facilitate this, all the recognized SNOMED-CT concepts are annotated in articles and can be added to the facet list. The adding and removal of concepts will trigger the data model to re-calculate the knowledge subgraph and graph partitioning given the edited facet list (\textbf{DG3}). 











%% file: tex/experiment.tex
\section{Experimental Evaluation}
We conducted a controlled user study to evaluate the usability of GRAFS and its ability to facilitate exploratory search and learning. We compared GRAFS to a typical faceted search interface with the same set of facets and documents to understand the effects of the newly introduced knowledge subgraph and visualization components. 

\subsection{Hypotheses}
\label{sec:Hypotheses}
The user study was designed to test the following hypotheses:

\textbf{Hypothesis 1.} With the help of the knowledge subgraph, GRAFS positively influences users' conceptual understanding and sensemaking activities.
\label{hyp1}
\begin{itemize}
    \item [\textbf{1.a}] We expect GRAFS to help users develop a better understanding of concept relations compared to the baseline system. The graphical \textit{facet list} and \textit{facet treemap} present relations between concepts in the current searching context and an overview of the large information space, which may otherwise require a substantial effort to build mentally via manual search and inference. 
    \item [\textbf{1.b}] Seeing concept relations helps users gain a deeper understanding of the search topic during sensemaking. Knowing concept relations should help users build a better conceptual understanding and mental model of the topic, and encourage users to explore more deeply into the results.
\end{itemize}

\textbf{Hypothesis 2.} The additional complexity introduced in GRAFS has no negative influence on users' searching activity. 
\label{hyp2}
\begin{itemize}
    \item [\textbf{2.a}] We expect that the introduction of the additional features in GRAFS will not hurt people's performance in facet filtering tasks. GRAFS preserved the layout and functionality of a faceted search interface, so users should be able to perform filtering tasks as in a typical faceted search interface. 
    \item [\textbf{2.b}] We expect the complexity of additional information provided in GRAFS will not hurt the overall usability of the system. The proposed data model and interface design present the concept relation information in a simplified way that facilitates users' learning and understanding of the complex information.

\end{itemize}




\subsection{Baseline Faceted Search Interface}

\begin{landscape}
\begin{figure}[t]
\centering  
\includegraphics[width=1\columnwidth]{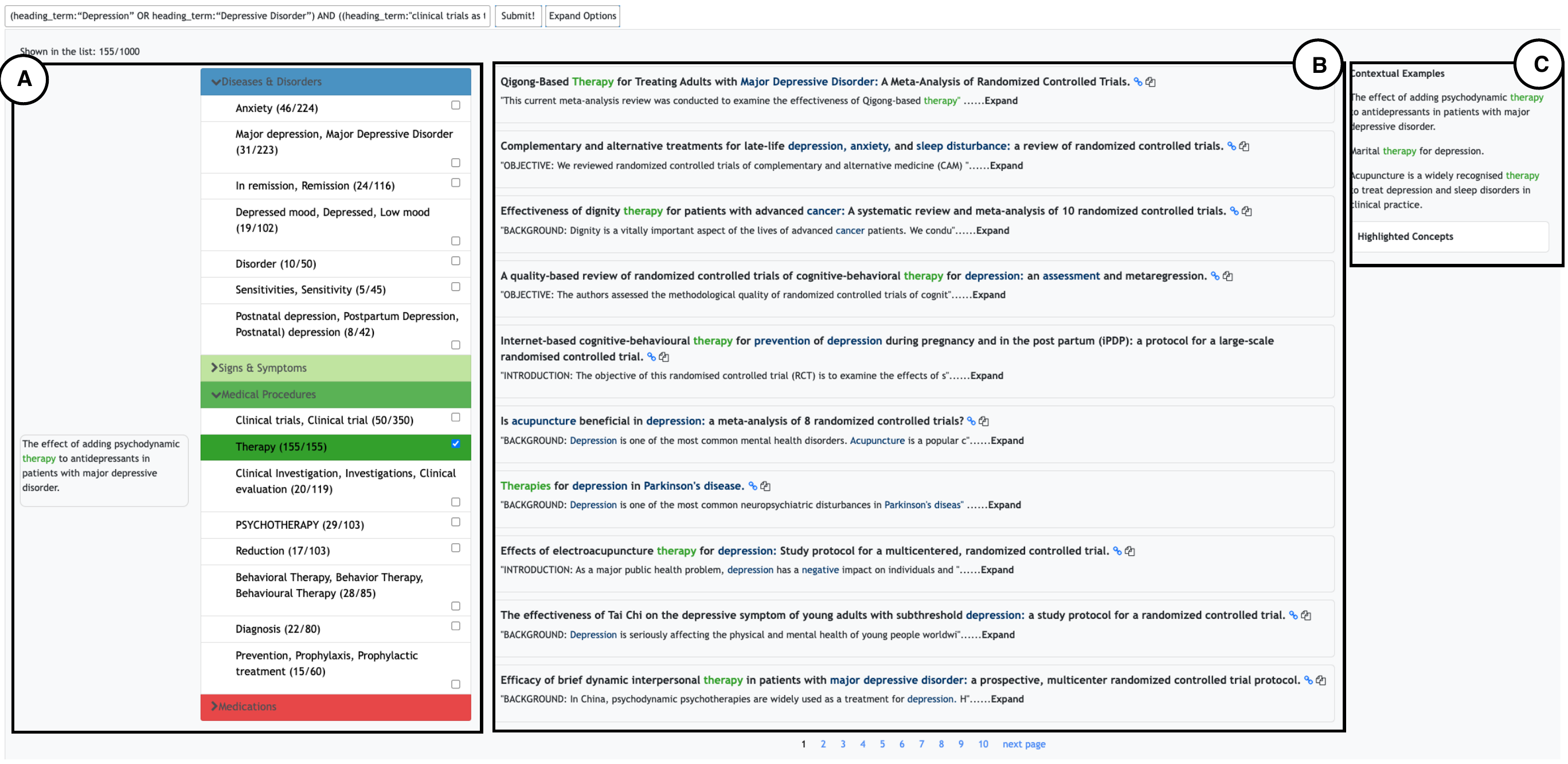}
\Description[Initial view]{Initial view}
\caption{Baseline search interface. It contains (A) \textbf{facet list} where concepts are organized by pre-defined categories, (B) \textbf{document list} showing filtered articles ranked by their relevance to the search query, (C) \textbf{concept provenance} that lists 3 examples of the selected concept.}
\label{fig:baseline}
\end{figure}
\end{landscape}

We compared GRAFS to a typical faceted search interface with a list of facets on the left and a ranked list of documents taking the major part of the page as shown in Figure \ref{fig:baseline}. To minimize the difference in the information conveyed by the two systems, we applied the same concept selection method as in GRAFS. In the baseline system, however, concepts are organized by pre-defined categories. The knowledge subgraph or graph partitioning are not surfaced in the baseline interface.

\subsection{Study Design}

\subsubsection{Overview}

We adopt a within-subjects design ($n=20$) where each participant was exposed to two Web-based search interface conditions, GRAFS and the baseline, working on two topics (``COVID-19 Diagnosis'' and ``Treatment for Depression''). As illustrated in Table \ref{tab:study_design}, 20 users were randomly split into four groups (i.e., five users per group). Each user completed two sessions. In each session, they worked on one topic using one system.

\begin{table}[H]
\begin{tabular}{|c|cc|cc|}
\hline
        & \multicolumn{2}{c|}{Session 1}             & \multicolumn{2}{c|}{Session 2}             \\ \hline
        & \multicolumn{1}{c|}{Topic}      & System   & \multicolumn{1}{c|}{Topic}      & System   \\ \hline
Group 1 & \multicolumn{1}{c|}{COVID-19}   & GRAFS    & \multicolumn{1}{c|}{Depression} & Baseline \\ \hline
Group 2 & \multicolumn{1}{c|}{Depression} & Baseline & \multicolumn{1}{c|}{COVID-19}   & GRAFS    \\ \hline
Group 3 & \multicolumn{1}{c|}{Depression} & GRAFS    & \multicolumn{1}{c|}{COVID-19}   & Baseline \\ \hline
Group 4 & \multicolumn{1}{c|}{COVID-19}   & Baseline & \multicolumn{1}{c|}{Depression} & GRAFS    \\ \hline
\end{tabular}
\caption{Experimental design. Participants were split into four groups. We counter-balanced the order of using the two systems and the topics used for testing the two systems.}
\label{tab:study_design}
\end{table}
To control for the difference between topics (e.g., topic difficulty and user familiarity), we counter-balanced the topics used for testing GRAFS and the Baseline. And to control for learning and fatigue effects, we counter-balanced the order of interface conditions. To control for differences in search results, we pre-specified search queries for each topic and did not allow participants to make edits to the queries. In principle, however, GRAFS does allow users to enter arbitrary search queries which are not pre-specified.

\subsubsection{Participants}
We recruited 20 participants (10 males, 10 females) via campus-wide mailing lists. All participants either were pursuing or had attained a graduate degree in a STEM field. Six participants had a biomedical background or professional experience, such as working as a Molecular Biology researcher. Nine participants' field of study is information science, four participants' field of study is computer science, and one participant's field of study is analysis and management.  Participants were randomly assigned to different conditions.

\subsubsection{Search Tasks}
\label{sec:search tasks}

Participants were asked to imagine being tasked to write a survey paper on either COVID-19 Diagnosis or Treatment for Depression. Given a search system and a query, such as GRAFS and COVID-19 Diagnosis, participants needed to complete two tasks: (1) produce an outline for the hypothetical paper (30 minutes), and (2) answer questions about specific aspects of the topic by facet filtering (10 minutes). 

\textbf{Task 1: Outline Generation}. 
To test \textbf{Hypothesis 1}, the user study task should be a realistic exploratory search task that requires conceptual understanding and sensemaking. Prior works on exploratory search and task design suggest the following desirable characteristics of exploratory search tasks: ``uncertainty, ambiguity, discovery, be an unfamiliar domain for the searcher, provide a low level of specificity about how to find the information, and be a situation that provides enough imaginative context for the participants to relate and apply the situation''~\cite{kules2009designing}. We designed our task to meet the above-mentioned characteristics. We adopted the scenario of developing an outline of a survey paper on biomedical topics. The process of generating an outline requires the user to make sense of the large information space of the search results, understand important concepts within the space, and organize pieces of information into a structure based on their understanding of how those pieces are related to each other. The task requires the user to quickly build a mental model of the search topic and present it in the format of an outline. At the same time, the included concepts and the structure of the participants-generated outlines surface their mental model of the topic and allow us to objectively evaluate the exploratory search outcomes.

Participants were asked to gather information and generate an outline of the hypothetical paper during the provided 30 minutes. Given the time limit, we suggested to participants that they should produce outlines in the form of bullet points with indentations that indicate the structure rather than organized, well-written paragraphs. Participants were instructed to provide sub-topics, major arguments or short descriptions for each sub-topic, and document references in their outlines to reflect the overall structure of the hypothetical paper. They can create and edit the outline at any point during the 30 minutes session. 


\textbf{Task 2: Question Answering.} The outline task evaluates users' exploration of a broad topic, which is often not how faceted search is primarily used. To test \textbf{Hypothesis 2.a}, we designed a question-driven task where participants focus on looking up answers for the following specific questions through facet filtering. 

\begin{itemize}
    \item \textit{COVID-19 Diagnosis:} Find evidence that antibody can be used to detect asymptomatic COVID-19 infection.
    \item \textit{Treatment for Depression:} Find if there are any clinical trials that study the effects of antidepressant medication.
\end{itemize}

Participants were given 10 minutes to find one or two articles that can answer the given question. They were not required to summarize the found articles.

\textbf{Rationale for Selecting the Search Topics:} As many of our participants are not from the biomedical domain, well-known health-related topics with relatively easy terminologies are favorable. Both COVID-19 and Depression are fairly familiar topics for laypeople, and they also involve professional concepts in biomedical literature that require learning and exploration. Therefore, we chose these two topics so that participants could start their exploration quickly and have the potential to learn unfamiliar concepts. 

\subsubsection{Outcome Measures}
\label{sec:outcome-measures}
We evaluated the system using both subjective feedback from users and an objective evaluation of the generated outlines. We first list the different types of measures used in the study and then connect the measures and metrics to each hypothesis.

\textbf{User Perceptions:} In the post-task questionnaire after each sub-session, participants provided their level of agreement with nine statements listed in Table \ref{table:post-task}, on a Likert scale from 1 (strongly disagree) to 7 (strongly agree). 

\textbf{Objective Evaluation of the Generated outlines:} We evaluated the quality of participant-generated outlines according to six rubric items as listed in Table \ref{table:outline}. Three graders were asked to select the better outline between the two generated by each user on all of the criteria, and ties are not allowed. The graders were not informed of the system used to generate the outlines and did the grading independently. In the end, we counted the total number of votes given to each system on each criterion.

\textbf{Time Spent on Task 2:} During the question-answering task, we measured the time spent by participants to find articles that they believed can exactly answer the given question.

\textbf{Qualitative User Feedback:} We asked for detailed feedback using a semi-structured exit interview with the questions in Table \ref{tab:interview}. We aimed to obtain open-ended findings besides testing our hypotheses regarding the general influence of GRAFS on exploratory search and the design insight of each interface component. 

\textbf{User Interactions}: As an auxiliary measure of qualitative user feedback, we counted the number of participants that used \textit{Facet List}, \textit{Facet Treemap}, \textit{Editable List}, and \textit{Concept Provenance} based on their action logs automatically collected during the user study. This piece of information helped us focus our qualitative analysis of a component's usability on the feedback provided by those participants who actually engaged with the component.

To keep a natural flow of user study, some measures (e.g., post-task questionnaire and exit interview) collect information to test multiple hypotheses. We highlight the connection between the measures and the hypotheses below. 

\textbf{Hypothesis 1}
\begin{itemize}
    \item \textit{User Perceptions.} In the post-task questionnaire, we listed four items (Q1-Q4 in Table \ref{table:post-task}) regarding how GRAFS helps the participants generate the outlines. Q2 aims to test \textbf{Hypothesis 1.a} by directly asking the participants whether GRAFS helped them learn concept relations. Q1, Q3, and Q4 are system qualities that can help users develop a complete and deep understanding of the topic, thus testing \textbf{Hypothesis 1.b}. 
    \item \textit{Objective Evaluation of the Generated Outlines.} We obtained objective quality evaluations of the outlines according to the six rubric items as listed in Table \ref{table:outline}. The six rubrics assess an outline's quality from different aspects. A1, A2, and A3 focus on the system's effect on sensemaking and learning, while A4 and A5 focus on the effect on people's searching and foraging. A6 measures the overall quality of the outline. If GRAFS helps participants gain a deeper understanding of the topic, participants should be able to generate outlines with higher quality when using GRAFS compared to using the baseline. We compared the number of votes given to GRAFS and the baseline to test \textbf{Hypothesis 1.b}. A1 also helps test \textbf{Hypothesis 1.a}, since the structure of the outline reflects a participant's understanding of how concepts are related to each other.
    \item \textit{Qualitative User Feedback of the system.} In the exit interview, we asked the participants to select the better outlines from the two they generated and a preferred system (I1 and I4 in Table \ref{tab:interview}). Participants' selections reflect the overall effectiveness of the system in the exploratory search task, thus helping test \textbf{Hypothesis 1}. We compared the number of votes given to GRAFS and the baseline. Furthermore, we did a qualitative analysis of participants' answers during the exit interview to gain insights into how GRAFS facilities the exploratory search process (Table \ref{tab:interview}).
\end{itemize}

\textbf{Hypothesis 2}
\begin{itemize}
    \item \textit{User Perceptions.} To test \textbf{Hypothesis 2.a}, we include one question in the post-task questionnaire regarding the easiness of doing the task (Q5 in Table \ref{table:post-task}). To test \textbf{Hypothesis 2.b}, we include four items (Q6 - Q9 in Table \ref{table:post-task}) related to the overall usability of the system. 
    \item \textit{Time Spent on Task 2.} The time spent on looking up answers during Task 2 measures the difficulty for the participants to conduct the facet filtering task using the system, thus helping test \textbf{Hypothesis 2.a}. 
    \item \textit{Qualitative User Feedback and User Interactions.} Participants' selections of the preferred system (I4) reflect the overall usability of the system, thus helping test \textbf{Hypothesis 2.b}. The qualitative analysis of participants' feedback I4 and I5 also helps test the usability of GRAFS and its individual components.
\end{itemize}
\begin{table}[h]
\begin{tabular}{llll}
\toprule
 & & Question & Hypothesis \\ \midrule
\multirow{3}{*}{\begin{tabular}[c]{@{}l@{}}Outline\\ generation\end{tabular}} & \textbf{Q1 Overview} & \begin{tabular}[c]{@{}l@{}}The faceted search tool helped me gain an\\ overview of key concepts in the search results.\end{tabular} & 1.b\\ \cmidrule{2-4} 
 & \textbf{Q2 Relation} & \begin{tabular}[c]{@{}l@{}}The faceted search tool helped me see how \\ key concepts are related in the search results.\end{tabular} & 1.a\\ \cmidrule{2-4} 
 & \textbf{Q3 Discover}  & \begin{tabular}[c]{@{}l@{}}The faceted search tool helped me discover\\ interesting aspects to explore further.\end{tabular} & 1.b\\ \cmidrule{2-4} 
 & \textbf{Q4 Decision} & \begin{tabular}[c]{@{}l@{}}The faceted search tool helped me decide which\\ aspects to focus on and which to ignore.\end{tabular} & 1.b \\ \midrule
\begin{tabular}[c]{@{}l@{}}Question \\ answering\end{tabular} & \textbf{Q5 QA difficulty} & \begin{tabular}[c]{@{}l@{}}I find it difficult to use the faceted search tool \\when I complete the question answering task.\end{tabular} & 2.a \\ \midrule
\multirow{4}{*}{Usability} & \textbf{Q6 Easy to use} & The faceted search tool is easy to use. & 2.b\\ \cline{2-4} 
 & \textbf{Q7 Manageable} & The faceted search tool is manageable. & 2.b \\ \cmidrule{2-4} 
 & \textbf{Q8 Stimulating} & The faceted search tool is stimulating. & 2.b \\ \cmidrule{2-4} 
 & \textbf{Q9 Well-organized}  & The faceted search tool is well-organized. & 2.b \\ \bottomrule
\end{tabular}
\caption{Post-task questionnaire.  Participants provided
their level of agreement on a Likert scale from 1 (strongly
disagree) to 7 (strongly agree). Each question is associated with a hypothesis.}
\label{table:post-task}
\end{table}

\begin{table}[h]
\begin{tabular}{llll}
\toprule
 &  & Description & Hypothesis\\ \midrule
\multirow{3}{*}{\begin{tabular}[c]{@{}l@{}}Sensemaking/\\ Learning\end{tabular}} & \textbf{A1 Structure} & \begin{tabular}[c]{@{}l@{}}Which outline is more logically \\organized/structured, as opposed to \\ a list of randomly ordered points?\end{tabular} & 1.a, 1.b \\ \cmidrule{2-4} 
 & \textbf{A2 Interpretation} & \begin{tabular}[c]{@{}l@{}}Which outline contains more \\interpretation written by users?\end{tabular} & 1.b \\ \cmidrule{2-4} 
 & \textbf{A3 Topical depth}& \begin{tabular}[c]{@{}l@{}}Which outline covers deeper  \\ issues of the topic, as opposed to \\superficial issues?\end{tabular}  & 1.b \\ \midrule
\multirow{2}{*}{\begin{tabular}[c]{@{}l@{}}Search/\\ Foraging\end{tabular}} & \textbf{A4 Documentation} & \begin{tabular}[c]{@{}l@{}}Which outline collects more papers \\ to  document their findings?\end{tabular} & 1.b \\ \cmidrule{2-4} 
 & \textbf{A5 Topical diversity} & \begin{tabular}[c]{@{}l@{}}Which outline covers more diverse \\ issues, as opposed to \\being narrowly focused?\end{tabular} & 1.b \\ \midrule
 & \textbf{A6 Overall} & \begin{tabular}[c]{@{}l@{}}Which outline is overall the better \\ one of the two?\end{tabular} & 1.b\\ \bottomrule
\end{tabular}
\caption{Outline assessment rubric. Each rubric item aims to test one or more hypotheses.}
\label{table:outline}
\end{table}

\begin{table}[h]
\begin{tabular}{lll}
\toprule
& Description & Hypothesis\\\midrule

\textbf{I1 Subjective outline evaluation}  & Which outline are you more satisfied with?  & 1.a, 1.b                                                                                                              \\\midrule
\textbf{I2 General strategy}               & \begin{tabular}[c]{@{}l@{}} Could you describe your strategy for \\ approaching the task?\end{tabular} & - \\\midrule
\textbf{I3 Benefit of each system}         & \begin{tabular}[c]{@{}l@{}}Did either of the faceted search tools help \\ you in any way to accomplish the task? \\ If so, how?  If not, why not?\end{tabular} & - \\\midrule
\textbf{I4 GRAFS vs baseline}             & \begin{tabular}[c]{@{}l@{}}If you were to do the same task again, \\ which of the two interfaces would you \\ prefer to use? Why?\end{tabular} & 1.a, 1.b, 2.b                  \\\midrule
\textbf{I5 Per-component usability}       & \begin{tabular}[c]{@{}l@{}}Did you find \{concept arcs, treemap, \\ example tooltip, deleting, adding\} useful? \\ Why? \end{tabular} & 2.b \\\bottomrule
\end{tabular}
\caption{Exit interview questions. Some of the questions aim to test one or more hypotheses. }
\label{tab:interview}
\end{table}


\subsubsection{Procedure}
User study sessions were conducted either in-person or remotely using Zoom video conferencing, and each lasted for roughly two hours. After providing informed consent, participants completed a demographics questionnaire that asked about their background and prior experience with literature search. A study session contained two sub-sessions that followed the same sequence of steps and a participant completed the outline task and question answering task on one topic using one interface design in each sub-session. At the beginning of a sub-session, we guided the participant through a hands-on tutorial of the search interface using an example topic. Participants were then informed of the topic to work on and asked to rate their level of familiarity with the topic on a scale from 1 (unfamiliar) to 7 (familiar). Then, participants completed the outline and question answering task. After the two tasks, participants finished a post-task questionnaire to provide feedback for the search interface. Finally, after completing the two sub-sessions, participants provided additional feedback about their experience using the tools through an exit interview.

\subsection{Data Analysis Methodology}
Below we describe the data analysis methods we applied to each outcome measure in Section \ref{sec:outcome-measures}.

\textbf{User Perceptions.} 
\begin{itemize}
    \item \textit{Analysis of system difference.}  The average ratings of the systems are compared to understand the difference between systems. 
    Fisher’s randomization test was used to test for significant effects due to system differences.
    \item \textit{Analysis of factors other than the system difference.} Participants' perceptions of both GRAFS and the baseline can be influenced by a variety of factors including their prior experiences and the order of using the systems. In our analysis, we split participants into groups considering the order in which they used the two systems ($G_{order}$), whether they have biomedical domain knowledge ($G_{domain}$), their experience with conducting a literature review ($G_{experience}$), and topic familiarity ($G_{familiar}$). We evaluated the effects of these group identities on the questionnaire responses with a mixed ANOVA model. The group identities were treated as between-subject factors and the system (GRAFS or baseline) was treated as a within-subject factor. We recorded the $F$ statistic, proportion of variance $\eta^2$, and $p$ value for each factor.
    \end{itemize}
    
\textbf{Objective Evaluation of the Generated Outline.} The number of votes given to the two systems by the graders is compared to understand the difference between the systems. 

\textbf{Time Spent on Task 2.} The averages of the time in seconds spent on Task 2 when using the two systems were compared. Fisher’s randomization test was used to test for significant effects due to system differences.

\textbf{Qualitative User Feedback.} 
\begin{itemize}
    \item \textit{Analysis of votes given to each system.} The number of votes given to the two systems by the participants when they answered I1 and I4 is compared to understand the perceived overall effectiveness and usability of the systems. 
    \item \textit{Qualitative analysis.} We conducted qualitative analyses on the detailed feedback regarding how the systems and system components influenced participants' activities and factors that influenced their performance on the tasks.
\end{itemize}

\textbf{User Interactions.} We analyzed the interaction logs of each user and computed whether they selected any concept in the facet list to view the arcs, hovered over any concept to view the concept provenance tooltip, selected any concept in the facet treemap, deleted any concept from the facet list or treemap, and added any concept from the search results.

%% file: tex/results.tex
\section{Results}


Overall, the results from our study support \textbf{Hypothesis 1.a} and \textbf{Hypothesis 2}. We observed less clear results for \textbf{Hypothesis 1.b}. Though there is some evidence supporting \textbf{Hypothesis 1.b}, our study also illustrates certain challenges in confirming the effects of different systems on exploratory tasks. This section presents the results and feedback obtained from our study and a number of insights uncovered during our analysis. Fisher's randomization test was used to test for  significant effects due to  system difference~\cite{smucker2007comparison}. We interpret an effect as \emph{significant} if $p < .05$, and \emph{weakly significant} if $.05 \le p < .1$.

\subsection{Overview}
\label{sec:result overview}
The presentation of the study's results and our key observations are organized according to the hypotheses defined in Section \ref{sec:Hypotheses}. Here we provide an overview of the main findings, with more details provided in the sections to follow.

\textbf{Hypothesis 1 Results}:  With the help of the knowledge subgraph, GRAFS positively influenced users' conceptual understanding and sensemaking activities. 

\begin{itemize}
    \item [\textbf{1.a}] We found that GRAFS made it easier for users to see the relations between concepts (Q2) (Sections \ref{sec:H1 User Perceptions}, \ref{sec:H1 Qualitative})
    \item [\textbf{1.b}] We received mixed results for Hypothesis 1.b. The objective evaluation of user-generated outlines showed that participants were able to generate outlines with more depth (A3) when using GRAFS, supporting our hypothesis (Section \ref{sec:Outline Task objective}).   The majority of participants also subjectively preferred to use GRAFS for the given tasks (I4), indicating that GRAFS facilitated participants' sensemaking activity (Section \ref{sec:H1 Qualitative}). This also supports the hypothesis. However, no difference was observed for the overall quality of user-generated outlines (A6) across the two systems, and participants did not indicate in the questionnaire (Q1, Q3, Q4) that GRAFS helped them develop a deeper understanding of the search topic (Sections \ref{sec:H1 User Perceptions}, \ref{sec:Outline Task objective}). Our results also showed that the baseline system helped participants collect more documents (A4) to support their arguments, and more participants believed the outlines they generated using the baseline is better (I1) (Sections \ref{sec:Outline Task objective}, \ref{sec:H1 Qualitative}).
\end{itemize}
\textbf{Hypothesis 2 Results}: The additional complexity introduced in GRAFS had no negative influence on users' searching activity.
\begin{itemize}
    \item [\textbf{2.a}] We found GRAFS did not make it more difficult to do the question-answering task evaluated subjectively by participants (Q5) (Section \ref{sec:H2 User Perceptions}) and objectively by the time participants spent on the question-answering task (Section \ref{sec:Question-Answering Task}).
    \item [\textbf{2.b}] We found GRAFS was rated similar to the baseline system on the usability metrics (Q6 - Q9) (Section \ref{sec:H2 User Perceptions}). GRAFS is more stimulating (Q8) and less manageable (Q7) than the baseline system (weakly significant). The majority of participants (14 out of 20) selected GRAFS for the future task (I4) (Section \ref{sec: H2 Qualitative}). Overall, participants provided positive feedback for the newly introduced components (I4, I5) (Section \ref{sec: H2 Qualitative}). However, the \textit{concept treemap} and the function of deleting and adding concepts were both more beneficial for users with a relatively deeper understanding of the topic compared to \textit{concept arcs} and \textit{concept example tooltips}. 
\end{itemize}

\begin{figure}[t]
\centering  
\begin{subfigure}{.48\textwidth}
    \centering
    \includegraphics[width=1\textwidth]{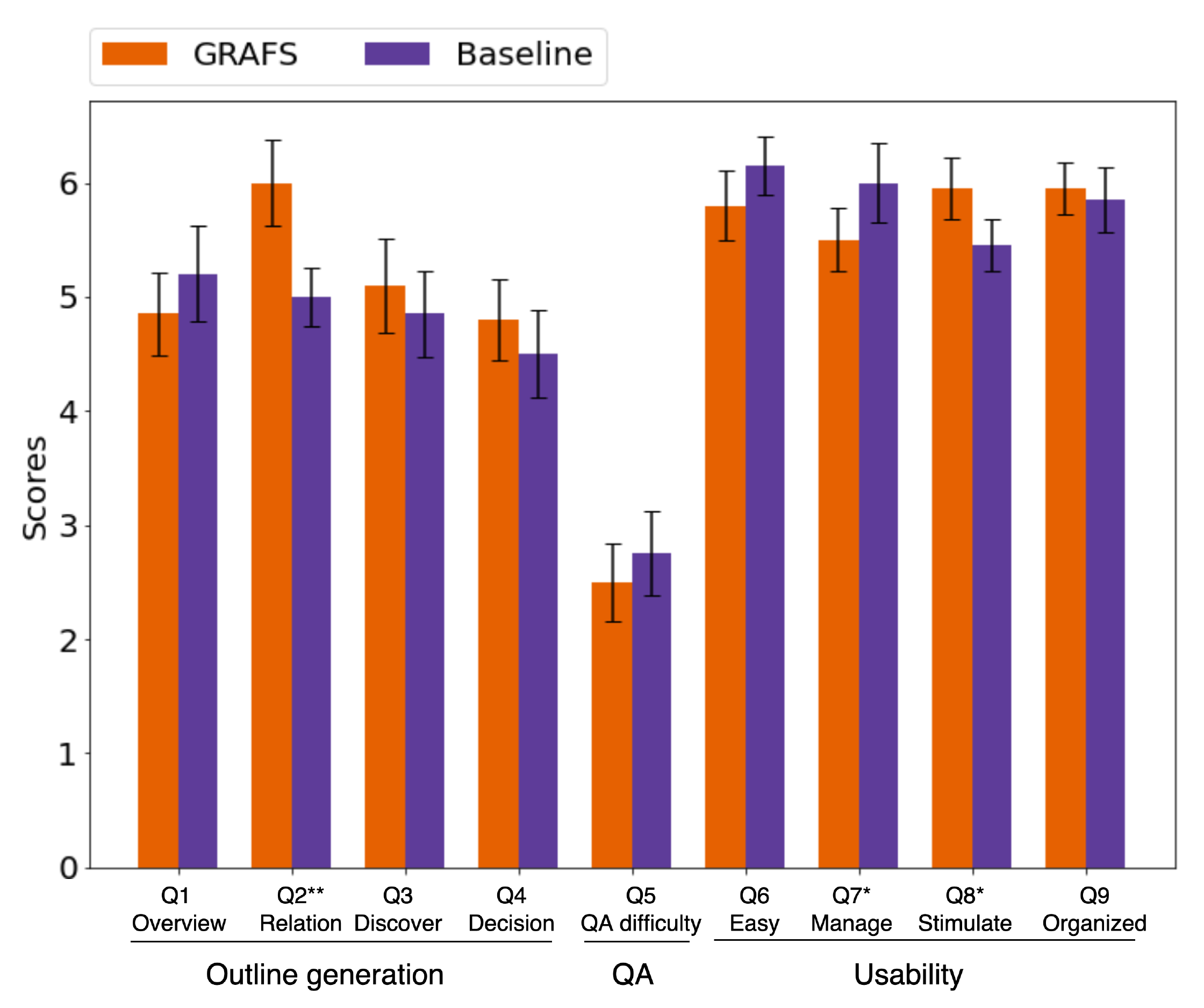}
    \caption{Post-task questionnaire results}\label{fig:res_a}
\end{subfigure}
    \hfill
\begin{subfigure}{.5\textwidth}
    \centering
    \includegraphics[width=1\textwidth]{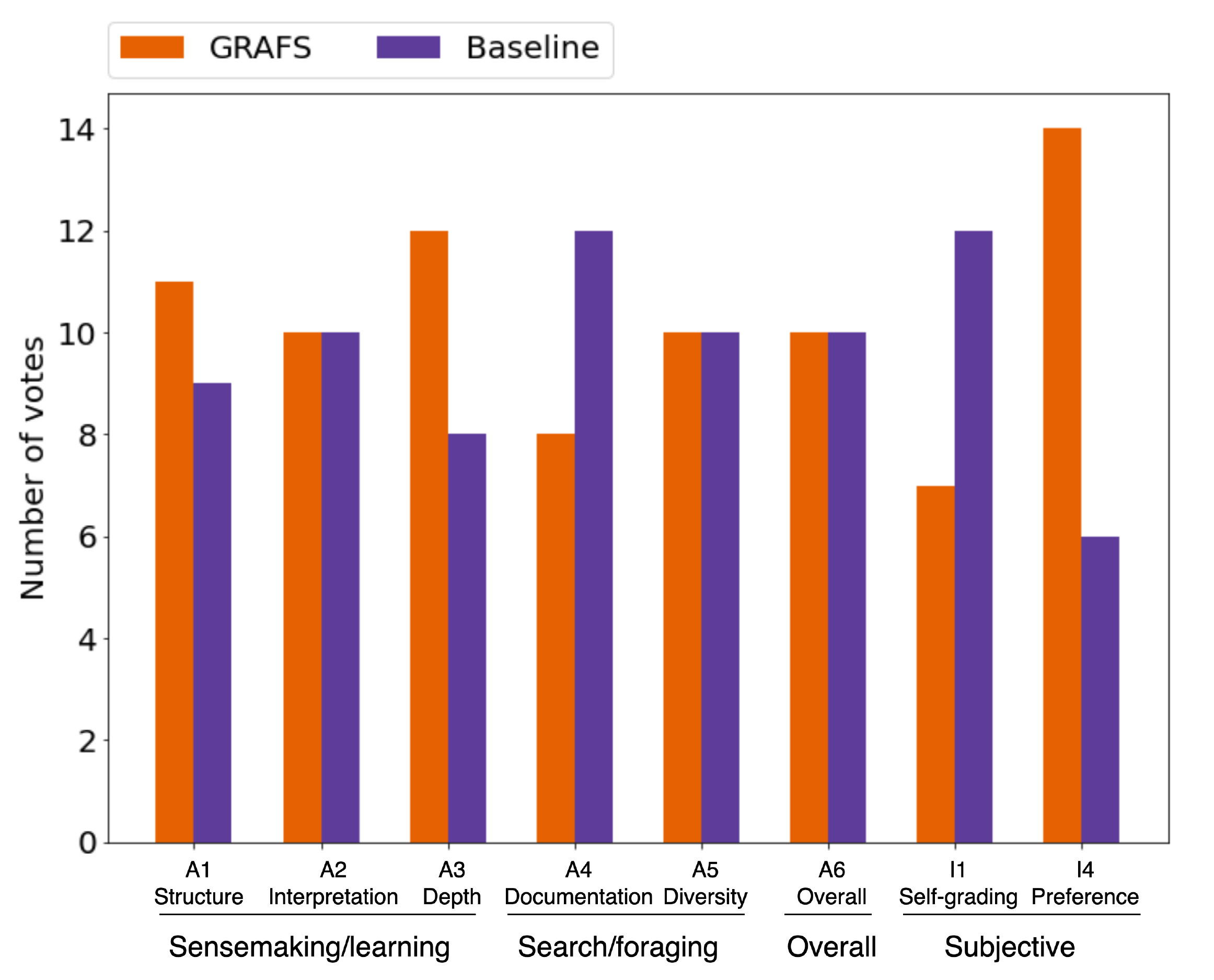}
    \caption{Outline generation task results}\label{fig:res_b}
\end{subfigure}
\caption{Results. (a) Mean ratings given by participants in the post-task questionnaire. Error bars show the standard error. There is a significant difference between GRAFS and the baseline in Q2 (**), and a weakly significant difference in Q7 and Q8 (*). (b) Numbers of votes given to each system by graders (A1-A6) and participants (I1, I4). GRAFS gets more votes in A3 Topical depth and I4 System preference, while the baseline gets more votes in A4 Documentation and I1 Self-grading. Overall, there is no difference between the outlines generated using the two systems (A6).}
\label{fig:res1}
\end{figure}

    

\subsection{Hypothesis 1}
Below we present detailed results regarding \textbf{Hypothesis 1} organized by relevant outcome measures.
\subsubsection{User Perceptions.}\label{sec:H1 User Perceptions}\hfill

\textbf{Analysis of system difference.} As shown in Figure \ref{fig:res1}(a), participants thought the two systems were different in helping them see the relation between concepts ($Avg_{GRAFS}=6.0$, $Avg_{Baseline}=5.0$, $p = 0.033$). This provides support for \textbf{Hypothesis 1.a} that GRAFS helps users see the relations between concepts better.

In contrast, the search system did not have significant effects on questions related to Q1 overview, Q3 discover, and Q4 decision. These results failed to show evidence supporting \textbf{Hypothesis 1.b}.

\begin{figure}[t]
\centering  
\begin{subfigure}{.32\textwidth}
    \centering
    \includegraphics[width=1\linewidth]{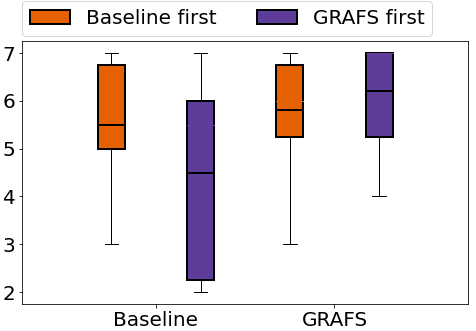}
    \caption{The effect of order on Q2 Relation}\label{fig:effect1}
\end{subfigure}
    \hfill
\begin{subfigure}{.32\textwidth}
    \centering
    \includegraphics[width=1\linewidth]{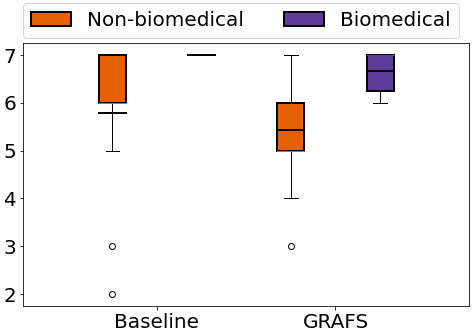}
    \caption{The effect of domain knowledge on Q6 Easy to use}\label{fig:effect2}
\end{subfigure}
    \hfill
\begin{subfigure}{.32\textwidth}
    \centering
    \includegraphics[width=1\linewidth]{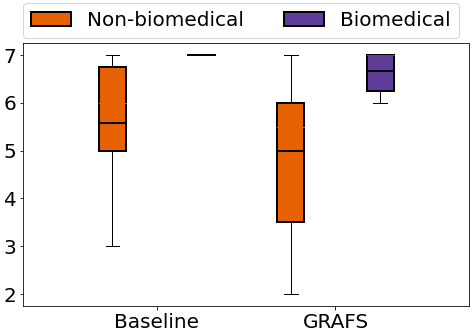}
    \caption{The effect of domain knowledge on Q7 Manageable}\label{fig:effect3}
\end{subfigure}
\caption{The effects of the order of using the two systems and participants' domain knowledge on participants' perception of the system. (a) Participants that worked on GRAFS first gave lower scores to the baseline in the system's ability to help them see the relation between concepts. (b) Biomedical domain experts gave higher scores to Q6 Easy to use compared to participants without domain knowledge. (c) Biomedical domain experts gave higher scores to Q7 Manageable compared to participants without domain knowledge.}
\label{fig:res_effect}
\end{figure}

\begin{table}[h]
\begin{tabular}{llll}
\hline
                    & $F$     & $\eta^2$ & $p$     \\ \hline
$G_{order}$        & 0.323 & 0.018  & 0.577 \\ 
$System$              & 6.406 & 0.262  & 0.021 \\ 
$G_{order}\times System$ & 3.139 & 0.148  & 0.093 \\ \hline
\end{tabular}
\caption{Effect of order on a user's rating of Q2 Relation}
\label{tab:order_effect}
\end{table}

\textbf{Analysis of factors other than the system difference.} Topic familiarity, literature search experience, and domain knowledge did not have any measurable effect on participants' responses on Q1-Q4. A weakly significant interaction effect between the order and system was found for participants' rating on Q2 Relation ($p = 0.093$, Table \ref{tab:order_effect}). As shown in Figure \ref{fig:res_effect}(a), when rating whether the system helps to see concept relations, users who worked on GRAFS first tended to give lower scores to the baseline system compared to the case when the user started from the baseline system. This phenomenon suggests that users did use some of the provided visual representations in GRAFS to perceive the relations between concepts. When exposed to the baseline system as the second interface, users appeared to have an increased awareness of the difficulty in seeing concept relations without the GRAFS features. This further supports our \textbf{Hypothesis 1}.

\subsubsection{Objective Evaluation of the Generated Outlines}
\label{sec:Outline Task objective}
Figure \ref{fig:res1} (b) shows graders' votes for the two systems (A1-A6). For A3 Topical depth, graders gave 12 votes to GRAFS and 8 to the baseline, supporting \textbf{Hypothesis 1.b}. In terms of A4 Documentation, 12 votes went to the baseline system and 8 went to GRAFS, indicating that being exposed to relation information may have negative effects on people's searching or foraging behavior. Therefore, A4 fails to support \textbf{Hypothesis 1.b}. One possible explanation is that when using GRAFS, participants may spend more time exploring concept relations and developing a deeper understanding, with less time left for collecting supporting documents. GRAFS and the baseline system received the same or the similar number of votes in A1 Structure, A2 Interpretation, A5 Topical diversity, and A6 Overall quality, which fails to support \textbf{Hypothesis 1.b}. Overall, A3 supports \textbf{Hypothesis 1.b} and the remaining criteria disagree with it in terms of the objective evaluation of outlines. 

\subsubsection{Qualitative User Feedback}\label{sec:H1 Qualitative}\hfill

\textbf{Analysis of Votes Given to each system.} Figure \ref{fig:res1} (b) shows participants' votes for the two systems (I1 and I4). When asked which system they would use if they were to do the task again, the majority of the participants ($n=14$) selected GRAFS (I4), supporting \textbf{Hypothesis 1}. When evaluating the generated outlines, a majority of the participants ($n=12$) felt that they did a better job producing their outlines using the baseline system (I1), failing to support  \textbf{Hypothesis 1}. 

Combined with questionnaire responses, one possible explanation is that being exposed to concept relations (Q2) increases complexity such that the user interface becomes less manageable (Q7, Section \ref{sec:H2 User Perceptions}). However, participants still found that relation information was useful for the exploratory task. Therefore, users expressed a preference for GRAFS because, if given more time on task, it would support a deeper understanding of the topic during exploratory literature search. In this way, participants' preference for GRAFS can be viewed as evidence for \textbf{Hypothesis 1}.

\textbf{Benefits of GRAFS and Baseline}
As mentioned previously, the majority of participants preferred GRAFS over the baseline system (14:6) when answering I4. During the exit interview, we asked participants to indicate the reason for their preference. In general, participants' feedback shows that the new information and functions provided by GRAFS facilitated users' exploratory activity while introducing some complexity to the use of the system. 

Participants described the benefits of GRAFS related to (1) relations, (2) specific components, and (3) general experience. Several users ($n=4$) mentioned that they loved the fact that they can learn relations between concepts in GRAFS (\textbf{Hypothesis 1.a}). Many participants ($n=7$) mentioned some components that they particularly liked, including facet treemap ($n=4$), editable list ($n=3$), and facet list arcs ($n=3$). Four participants selected GRAFS for a better overall experience. Specifically, they mentioned GRAFS is interesting and colorful ($n=2$), and helps with overview ($n=1$) and new ideas ($n=1$). 

Most of the participants that preferred the baseline system thought the baseline system is clearer ($n=4$), while two mentioned that they liked the pre-defined categories ($n=2$).

\textbf{Participants' Feedback on Factors Affecting the Generated Outlines.} During the exit interviews, participants provided detailed reasons for why they preferred the outline they generated with one system over the other when answering I1.  Participants' comments indicate that the interface design did affect their activity when completing the task, but the influence of the search topic on a user's outline proved to be larger. This may help explain why we did not observe major differences in the quality of people's outlines in A1, A2, A5, and A6 (Section \ref{sec:Outline Task objective}).
 
Reflecting this, most of the participants ($n = 12$) mentioned the influence of topic difference on the outline quality. This includes the technicality of a topic, participants' familiarity with a topic, and whether a topic is evolving or well-established. Some participants ($n = 6$) thought ``Treatment for Depression'' was easier than ``COVID-19 Diagnosis'' in terms of vocabulary and the topic's structure. Five participants mentioned that they were more familiar with one of the topics, and those participants felt that they did a better job on the familiar topic. Some participants ($n=4$) mentioned that the two topics were different from each other in the work required to perform the task, which in turn influenced the quality of the final outcomes. Three out of the four participants commented that they found depression to be a well-established field while COVID-19 related research was newer (and therefore harder to approach due to the more scattered and heterogeneous set of related concepts).


A number of participants ($n=6$) mentioned that the features of the interface affected the quality of their outline. Four users said that GRAFS helped them generate good summaries as it enabled them to edit the concept list ($n=2$), see relations between concepts ($n=1$), and was easier to use ($n=1$). In contrast, one participant thought the features in GRAFS were distracting and negatively influenced his work. One participant liked the categories given in the baseline system and reported that they helped her to write a good summary.

Overall, we found that topic differences and participants’ backgrounds may have a bigger impact on outline quality than the system. On the one hand, these factors may exaggerate the number of votes given to one system versus another due to the small sample size of the study. On the other hand, these factors may also obscure system differences (i.e., if participants always perform better on one topic, then GRAFS and Baseline will always get equal votes).

\subsection{Hypothesis 2}
Below we present detailed results regarding \textbf{Hypothesis 2} organized by relevant outcome measures.

\subsubsection{User Perceptions.}
\label{sec:H2 User Perceptions}
\hfill

\textbf{Analysis of system difference.} As shown in Figure \ref{fig:res1}(a), the questionnaire results generally support \textbf{Hypothesis 2}. There is no significant difference in Q5, indicating that the additional features in GRAFS did not make the question answering task more difficult (\textbf{Hypothesis 2.a}). The system usability questions (Q6-Q9) provide supporting evidence for \textbf{Hypothesis 2.b}. There is no significant difference in Q6 Easy to use and Q9 Well-organized, while the differences in Q7 Manageable ($Avg_{GRAFS}=5.5$, $Avg_{Baseline}=6.0$, $p = 0.099$) and Q8 Stimulating ($Avg_{GRAFS}=5.95$, $Avg_{Baseline}=5.45$, $p = 0.092$) are only weakly significant.

\begin{table}[h]
\begin{tabular}{llll}
\hline
                    & $F$     & $\eta^2$ & $p$     \\ \hline
$G_{domain}$       & 6.170 & 0.255  & 0.023 \\ 
$System$              & 1.797 & 0.091  & 0.197 \\ 
$G_{domain}\times System$ & 0.002 & 0.000  & 0.967 \\ \hline
\end{tabular}
\caption{Effect of domain knowledge on a user's rating of Q6 Easy to use}
\label{tab:knowledge_effect1}
\end{table}

\begin{table}[h]
\begin{tabular}{llll}
\hline
                    & $F$     & $\eta^2$ & $p$     \\ \hline
$G_{domain}$       & 8.160 & 0.312  & 0.010 \\ 
$System$              & 3.635 & 0.168  & 0.073 \\ 
$G_{domain}\times System$ & 0.173 & 0.010  & 0.682 \\ \hline
\end{tabular}
\caption{Effect of domain knowledge on a user's rating of Q7 Manageable}
\label{tab:knowledge_effect2}
\end{table}

\textbf{Analysis of factors other than the system difference.} Participants with biomedical background gave significantly higher scores for Q6 Easy to use ($p = 0.023$, Table \ref{tab:knowledge_effect1}, Figure \ref{fig:res_effect}(b)) and Q7 Manageable ($p = 0.010$, Table \ref{tab:knowledge_effect2}, Figure \ref{fig:res_effect}(c)) for both GRAFS and the baseline system. The effect of $G_{domain}$ is more significant than the effect of the system on the responses for these two questions. This suggests that people's judgment of a system's usability is highly influenced by people's domain knowledge. Topic familiarity, literature search experience, and the order of using the two systems did not have any measurable effect on participants' responses on Q5-Q9.

\subsubsection{Time Spent on Task 2.} 
\label{sec:Question-Answering Task}
Participants indicated their perception of the difficulty of the question answering task through the post-task questionnaire. As mentioned in Section \ref{sec:result overview}, there is no significant difference between the ratings given to the two systems regarding the task's difficulty ($p=0.510$). We further evaluated participants' performance based on the time spent on the task. On average, participants spent 181.1s on the question answering task using GRAFS and 238.2s using the baseline system, and the difference is not significant ($p=0.208$). This shows that participants performed the question answering task at least as well on GRAFS as on a typical faceted search system, which supports \textbf{Hypothesis 2.a}.

\subsubsection{Qualitative User Feedback.} 
\label{sec: H2 Qualitative}\hfill

\textbf{Analysis of Votes Given to each system.} When asked which system they would use if they were to do the task again, the majority of the participants ($n=14$) selected GRAFS (Figure \ref{fig:res1} (b), I4), implying that participants perceived GRAFS to be usable overall. This result thus helps support \textbf{Hypothesis 2.b}.

\textbf{Benefits of GRAFS and Baseline.} During the exit interview, we asked participants to describe why they prefer one system over the other. Participants described the benefits of GRAFS related to (1) relations, (2) specific components, and (3) general experience. The detailed comments are presented in Section \ref{sec:H1 Qualitative}. These comments reflect that a majority of the participants found the newly introduced components usable and were able to apply that new information to their tasks, providing support for \textbf{Hypothesis 2.b}. However, some participants also pointed out that the baseline system is clearer ($n=4$), implying that the newly introduced features in GRAFS add complexity to the use of the system.

\textbf{Usability of Interface Components.} During the exit interview, participants were asked whether a component is useful or not for accomplishing the given tasks.  We note that there is a difference between (1) a participant not using a feature, and (2) a participant finding the feature not useful after using it. More specifically, we noticed that some participants did not use certain features in GRAFS due to time limitations, lack of domain knowledge, and limited experience with a new interface design.
Therefore, we report both (1) the number of participants that used each feature based on the action log, and (2) the number of participants that found each feature to be useful. 

Overall, the facet list arcs were the most used feature. The facet treemap was used less often, in part due to its higher complexity. The functions of deleting and adding concepts were the least used. Though they all add complexity to the traditional faceted search interface, the components all received positive feedback from participants who have used them. The concept provenance tooltip was found to be easy to use, but not as useful as other components.

\textbf{Facet List Arcs:} Action logs show that all 20 participants made at least one selection from the facet list, which triggers the display of facet list arcs. The majority of participants thought the feature of concept arcs to be useful ($n=11$). Some participants ($n=6$) mentioned that arcs helped inform them about what to look at or which concepts to click next (\textbf{Hypothesis 1.b}). Five participants mentioned that arcs helped them understand relations between concepts (\textbf{Hypothesis 1.a}), which in turn helped them better understand the topic and adjust their selection strategy (\textbf{Hypothesis 1.b}).

For participants that did not state that arcs were useful, a majority of them said that they ignored this feature ($n=5$). For instance, User 4 said that he preferred to just read articles and User 12 mentioned that due to time limitations, he was not able to use that information. Some participants ($n=2$) used alternatives to the arcs to get similar information, such as the facet treemap, or the number of documents listed near each concept.

\textbf{Facet Treemap:} We considered a participant to have used the facet treemap if they clicked at least once on a concept in the visualization. 13 participants meet this criterion. Participant feedback illustrates that the facet treemap can be difficult to make sense of due to its unfamiliarity, and it introduces complexity to using the system. Yet while it may take time for users to learn how to use this feature, nine out of 20 participants thought that the facet treemap was useful. 

Most of the participants who found it useful ($n=8$) reported that the facet treemap helped them see the relations between concepts (\textbf{Hypothesis 1.a}). Because both the facet treemap and the facet list arcs communicate the connections between concepts, some participants ($n=7$) compared the two. Five participants indicated that they preferred facet treemap compared to facet list arcs, while the other two participants preferred arcs. Those two felt that the arcs were more straightforward. 

Interestingly, among people that preferred the facet treemap, two participants commented that the facet treemap was not straightforward to understand and that it took time for them to find it useful. For instance, User 20 mentioned that he didn’t find the treemap to be useful immediately. However, during the exploration, he found that thick arcs in the facet list always seemed to connect to broader concepts such as infectious disease. He, therefore, started to look to the treemap for a more nuanced view of the relations and, eventually, found the treemap to be more useful. User 14 mentioned that at the beginning of the session, she felt the information in the facet treemap is cluttered and the structure seemed to be quite complex. As a result, she assumed that it would be hard to understand the treemap and ignored it. Only after using the facet treemap did she realize that it is easy to use.


A majority of the participants ($n=11$) did not report the facet treemap as being useful. Three participants felt the information was cluttered and overwhelming, while some participants ($n=2$) thought it occupied too much screen space. Three participants mentioned that it was difficult to make sense of the information provided by the facet treemap, and two participants said they were not used to information presented in that way. Many of these comments suggest unfamiliarity as a key hurdle.

\textbf{Concept Provenance Tooltip:} All participants used this feature when hovering over items in the facet list. Eight participants mentioned that the facet list tooltip was useful. Some participants ($n=4$) mentioned that the tooltip helped them gain a quick understanding of what the concept meant. Two participants expressed the wish that they could navigate directly to an original article from the given examples. 

For people that did not report that the facet list tooltip was useful, four mentioned that they did not need extra explanations of the concepts either because they had good background knowledge or they only used familiar concepts during their exploration. Some participants ($n=3$) preferred to understand concepts using other methods, such as ``googling'' or reading the actual articles. One participant mentioned that the usefulness of the tooltip was highly dependent on the content of the sentence shown in the tooltip. Overall, the concept provenance tooltips were found to help people understand concepts during exploration, but only when users were faced with new concepts.

\textbf{Editable List:} GRAFS allows users to make adjustments to the automatically generated knowledge subgraph by deleting or adding concepts. Overall, this feature was less used by participants due to task time  constraints, limited familiarity with the topic, and limited training with this way of interacting. However, participants provided positive feedback on this feature, and we found some interesting use cases that indicate the benefits of preserving human agency in GRAFS.

In total, seven participants used the function of either deleting or adding concepts. Out of these seven, four participants used both, two participants only deleted concepts, and one participant only added concepts.

We examine the feedback regarding deleting concepts and adding concepts separately. Six of the 20 participants used the function of deleting concepts, and six participants (four of the six participants that used concept deletion, plus two who did not use concept deletion) mentioned that the function of deleting concepts was useful for removing irrelevant or equivalent variants of concepts. This helped users focus on a shortened concept list ($n=2$) and find important relations faster ($n=1$). 

Five of the 20 participants used the function of adding concepts, and four participants (three of the five participants that used concept addition, plus one who did not use concept addition) found adding concepts to be useful. Two people used this function when they came across unlisted concepts that they felt were important based on their reading. An interesting use case was provided by User 2. She described that when she worked on ``Treatment for Depression,'' she frequently came across the concept of ``Exercise.'' Therefore, she added ``Exercise'' to the concept list to investigate deeper. When she felt she had done enough reading of related materials, she then removed the concept from the list. One participant added the concept ``COVID-19,'' which was removed as a per-query stop word, to remind herself about the major topic. One participant suggested the interface should enable users to add new concepts to the facet list by directly typing in concept names in addition to selecting concepts in documents.

For participants that did not use the add or remove functions, four participants mentioned that time limitation was a major reason. Some participants ($n=3$) mentioned that the lack of familiarity with the topic is another important factor.

%% file: tex/discussion.tex
\section{Design Implications}
\label{sec: Design Implications}
\subsection{The Effect of Showing Concept Relationships}

The major difference between GRAFS and the baseline faceted search interface is the focus on revealing relationship information between key concepts. The results from our experiments show that this additional information can positively contribute to a user's conceptual understanding of a search topic. Showing these relationships as arcs nudges users to build more a sophisticated mental model of the information space as reflected by conceptually deeper outlines. Notably, these benefits were achieved without substantially impacting system usability. In user experience terms, the relationship arcs serve as \emph{signifiers} that make conceptual relationships more \emph{discoverable} by users~\cite{norman2013design}. These signifiers may not be needed in relatively simple exploratory tasks such as comparative shopping, but can be particularly valuable in research-oriented tasks where users need to explore, discover, and learn about an unfamiliar and complex information space. A future research direction is to characterize task scenarios where it is most beneficial concept relationships to augment a faceted search interface.

The mixed results for Hypothesis 1.b (``seeing relationships helps users gain a deeper understanding of the search topic'') indicate that objective and subjective evaluation of learning outcomes may not  align. Although GRAFS helped participants construct objectively deeper and more organized outlines, more participants favored the outlines generated using the baseline system that had a simpler logical structure but contained more papers. Such results imply a tension between \emph{learning} and \emph{satisfaction} in exploratory search. Learning activities on GRAFS may have not only slowed participants down in terms of getting more papers into their outline, but also exposed them to a larger sphere of knowledge and made them think their outline was not thorough enough.  In other words,  learning made users aware of what they do not know and feel less satisfied with what they have already known. In psychology research, studies also observed that people's self-evaluations of test results diverge further from objective evaluations when they are less knowledgeable on the test subject~\cite{kruger1999unskilled}. 
The mixed results for Hypothesis 1.b have implications for future research in two aspects. First, information retrieval systems that support learning and sensemaking should explore methods for computationally estimating users' learning progress so that they can inform users of their progress. Initial work in this direction is recently explored in search-as-learning literature~\cite{urgo2022learning}. Second, such systems' interface should communicate  user's learning progress in a positive tone and encourage them to explore further with a sense of achievement. An inspiring line of related work is in online news consumption diversification, where the goal is to encourage users to discover and read  news from diverse political viewpoints~\cite{munson2013encouraging}.

\subsection{The Value of User Agency}

Our study suggests that it is important to preserve human agency when the system takes the initiative to recommend an imperfect data model. In GRAFS, we allow users to adjust the automatically generated data model by deleting and adding concepts. Though this feature was less used in the study, it is one of the most mentioned components when participants talked about the benefits of GRAFS. The use cases provided by participants show that they adjust the concept list based on their exploratory focus, such as concepts they want to ignore, keep track of, and look deeper into. We believe this feature will be more widely used when users have more time in actual exploratory search tasks, as many participants mentioned that they did not edit the concept list because of either time limitations or lack of familiarity with the topic.

To increase user agency, a future improvement opportunity is to allow users to directly 
add concepts to the facet list by typing the concept name, and the system can assist the user through auto-completion. This would allow users who have prior knowledge to directly specify concepts they are interested in without having to find them in the result list. Even for users who learned a new concept on the task, such a feature would allow them to add a newly learned concept by recalling it from memory without having to refind a document mentioning the concept and then visually search for the concept inside the document.

\subsection{When Less Is (Not) More In Surfacing Complex Information}
Many design elements in GRAFS aim to minimize information overload, reflecting the minimalist motto ``\emph{less is more}.’’ First, the original query-specific knowledge graph containing hundreds of concepts is reduced to a much smaller initial subgraph with 20 or so key concepts. Second, among all relationships in the smaller subgraph, only five arcs connecting the most related concepts are shown at any time. Third, the original densely connected subgraph is reduced to a tree structure (facet treemap). The facet list further flattens this tree structure into a list and is better received by our participants than the tree. All of these elements selectively expose users to a small but informative portion of the underlying information, balancing complexity and usability. In the case of relationship arcs, the exposure is both selective and progressive -- new arcs are gradually surfaced as the user explores different combinations of concepts.

However, not all data reductions were perceived as useful by participants. For example, our concept provenance features (hover-over tooltip on the facet list and a textbox on the right side of the screen) were designed to explain why a concept was considered relevant given the search context. The features showed example sentences (instead of whole documents) that mentioned both the concept and the current query. Participants’ qualitative feedback suggested that the provided sentences were not helpful, or even viewed as a distraction for some participants who already had a good background. 
In addition, some participants preferred more comprehensive explanations from a separate search or a larger context, such as the original articles. As the need for explanation varies between users and contexts, a possible solution is to only show concept provenance upon request, and to ease the navigation from provenance sentences to their original documents.

These phenomena imply that showing a small part of a larger data structure can help reduce information overload \emph{if that data structure is self-similar}. For example, a subgraph is structurally similar to the larger knowledge graph; a subset of relationship arcs is structurally similar to the set of all arcs; a low-dimensional projection of a network is structurally similar to the original network representation. In these cases, users can still make sense of the smaller part as it preserves the ``syntax'' of the whole.
However, fragments of natural language data (concepts and sentences) are integral parts of larger contexts (documents). Showing them out of context may result in ``information underload'' due to syntactic incompleteness, semantic ambiguity, and lack of coherence. This is especially true in scenarios where users are unfamiliar with the information space and may encounter difficulty in interpreting extracted concepts and sentences without seeing the original context. Therefore, ``less is more'' applies in some information presentation scenarios but not others. 

\section{Limitations and Challenges}
\label{sec:Limitations and Challenges}
\subsection{Limitations of the System}
The proposed data model and visualization elements have two key limitations: (a) they introduce complexity to users' search when trying to present richer information; and (b) they are less familiar to users compared to a typical faceted search interface. To cope with these limitations, we built our system on top of a typical faceted search interface and only added visualizations to the marginal areas of the interface. We attempted to preserve the simplicity and familiarity of the system, and gave users the freedom to ignore additional components. Reflecting this we observed participants that used GRAFS as a Google-like search engine or who ignored new parts of the interface such as the arcs and treemap. Our study suggests that the added interface complexity in GRAFS did not interfere with basic information retrieval tasks (e.g., looking up specific articles). In terms of subjective perception, participants indicated that GRAFS was less manageable compared to a typical faceted search interface. 
The same users rated GRAFS and the baseline similarly in ``easy to use.'' 
In retrospection, a potential solution would be to allow users to manage the visual complexity by turning certain features ``on'' or ``off'' (e.g., through toggle switches).

\subsection{Limitations of the User Study}
Two main limitations of the user study were (a) the large impact of the search topic and participants' background; and (b) the time constraint. First, participants' background and differences between topics may have a much larger influence on the study outcomes than the system, especially for a challenging task involving searching, reading, and writing skills. A within-subjects design may not solve this problem because even the same participant may have different background knowledge in different topics. Second, it takes time for users to make sense of the visualized information. Under the time pressure of a formal study session, participants might be discouraged to use new tools and instead follow their familiar approach to doing the task. Even when participants were willing to use the visualization components, it also required time and effort to fit the new information these new tools provided into their traditional workflow. In future work, it would be valuable to study the system's usability by observing people using the system as they work on longer-term real-world tasks.

Two other aspects of the study design could serve as potential limiting factors. First, due to the timing of this study which overlapped with the COVID-19 pandemic, some study participants took part in the study virtually via Zoom while others participated in a face-to-face session. While we did not identify any specific impacts of participant modality in our analysis of the results, the mode of participation was a potential confounder in the execution of the study.  Second, the post-task questionnaire (see Table~\ref{table:post-task}) included a series of agree/disagree Likert scale questions. The phrasing of the questions as agree/disagree questions could potentially lead participants to record more positive ratings than they would provide if asked for similar feedback using alternative question formats.

%% file: tex/conclusion.tex
\section{Conclusion}

This paper proposed graphical faceted search (GRAFS), a novel interactive approach designed to help exploratory users more effectively construct a mental model of an unfamiliar information space during exploratory search. 
GRAFS leverages an intelligent backend computational model that extracts a small but essential set of key concepts and their relations present in semantic search results to form a knowledge subgraph.
The frontend search interface leverages this subgraph to organize and visualize search information in a faceted search-like interface that users are familiar with. Users are guided by the computed subgraph, which is itself updated based on user search activity. 
We conducted a user study that compared the proposed GRAFS approach against a baseline faceted search system in the context of exploratory literature search. Experimental results show that the proposed approach can effectively help users recognize relationships between key concepts, leading to a more sophisticated understanding of the search topic while maintaining  similar functionality and usability as a classical faceted search system.